\documentclass[prb,twocolumn,amsmath,amssymb,superscriptaddress,longbibliography]{revtex4-2}

\usepackage[T1]{fontenc}
\usepackage[utf8]{inputenc}
\usepackage[english]{babel}

\usepackage{amsmath,amssymb,amsfonts,bm}
\usepackage{mathtools}
\usepackage{braket}
\usepackage{chemformula}

\usepackage{graphicx}
\usepackage{xcolor}
\usepackage{color}
\usepackage{tikz-feynman}
\usepackage{array}
\usepackage{longtable}

\graphicspath{{figures_prl/}}

\usepackage{enumitem}
\usepackage[
    breaklinks,
    colorlinks,
    bookmarks=false,
    citecolor=blue,
    linkcolor=red,
    urlcolor=blue
]{hyperref}

\hypersetup{
    colorlinks,
    allcolors=blue
}

\usepackage{cleveref}
\usepackage{orcidlink}

\usepackage{etoolbox}

\crefname{equation}{Eq.}{Eqs.}
\Crefname{equation}{Eq.}{Eqs.}
\crefname{figure}{Fig.}{Figs.}
\Crefname{figure}{Fig.}{Figs.}
\crefname{section}{Sec.}{Secs.}
\Crefname{section}{Sec.}{Secs.}
\crefname{table}{Table}{Tables}
\Crefname{table}{Table}{Tables}

\newcommand{\nn}{nearest-neighbor}

\newcommand{\fm}{ferromagnetic}

\newcommand{\canted}{c-120$^{\circ}$}

\makeatletter
\appto{\appendix}{%
  \@ifstar{\def\theequation@prefix{A.}}%
          {}%
}
\makeatother

\begin{document}

\makeatletter
\renewcommand{\fnum@figure}{FIG.~\thefigure}
\renewcommand{\fnum@table}{TAB.~\thetable}
\makeatother

\title{Incommensurate Spin-Density Waves in a Frustrated Maple-Leaf Lattice Ferromagnet}

\author{Paul L. Ebert\,\orcidlink{0000-0003-1614-6920}}
\email{pebert@pks.mpg.de} 
\affiliation{Max Planck Institute for the Physics of Complex Systems, N\"othnitzer Strasse 38, Dresden 01187, Germany}

\author{Yasir Iqbal\,\orcidlink{https://orcid.org/0000-0002-3387-0120}}
\email{yiqbal@physics.iitm.ac.in}
\affiliation{Department of Physics, Indian Institute of Technology Madras, Chennai 600036, India}

\author{Alexander Wietek\orcidlink{0000-0002-4367-3438}}
\email{awietek@pks.mpg.de}
\affiliation{Max Planck Institute for the Physics of Complex Systems, N\"othnitzer Strasse 38, Dresden 01187, Germany}
\affiliation{Department of Physics, Indian Institute of Technology Madras, Chennai 600036, India}


\begin{abstract}
We study how ferromagnetism breaks down in the spin-$\tfrac12$ nearest-neighbor Heisenberg model on the maple-leaf lattice with ferromagnetic $J_t,J_d$ and antiferromagnetic $J_h$, motivated by the mixed ferro-antiferromagnetic interactions in Na$_2$Mn$_3$O$_7$.
Exact diagonalization shows that the ferromagnetic boundary does not feature a zero-field spin-nematic phase on the clusters studied here, but an extended regime of incommensurate spin-density-wave correlations with continuously evolving ordering vector. The phase diagram also contains collinear N\'eel, canted $120^\circ$, and hexagonal-singlet regimes, separated by regions that remain difficult to classify from exact diagonalization alone. Variational tests of fully symmetric Gutzwiller-projected Abrikosov-fermion U(1) and $\mathbb{Z}_2$ states find no competitive spin-liquid description of the interior unresolved regions. By contrast, on the ruby-lattice boundary we identify a point between the collinear N\'eel and hexagonal-singlet phases where a projected $\mathbb{Z}_2$ Ansatz reproduces the finite-size energy and spin correlations with good accuracy.
\end{abstract}

\date{\today}

\maketitle

\paragraph*{Introduction.} 

\begin{figure}[t]
        \centering
        \includegraphics[width=0.9\linewidth]{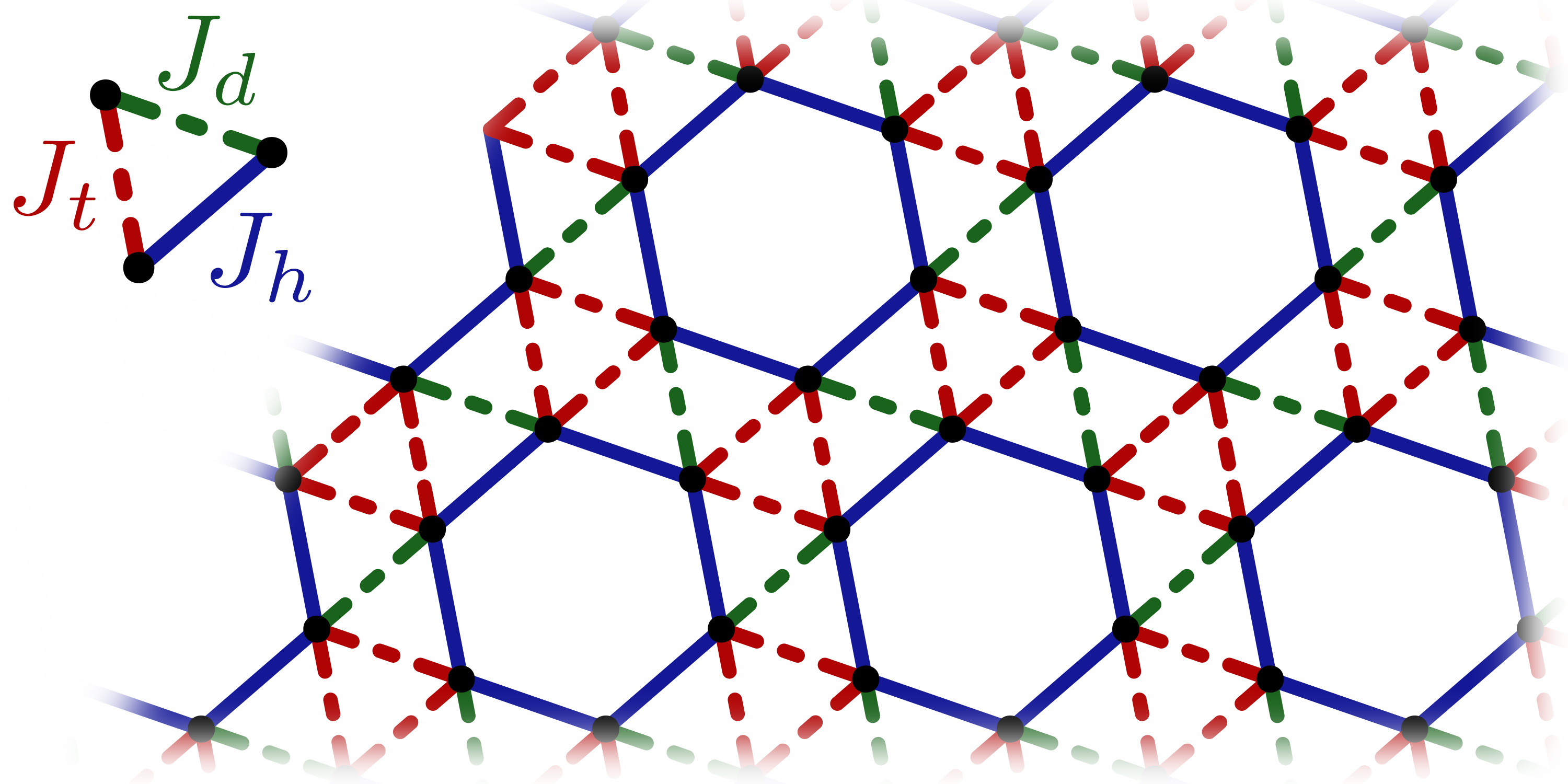}
        \caption{\label{fig:lattice}
            Maple-leaf lattice with three \nn{} bonds: $J_t$ (red triangles), $J_d$ (green ``dimers''), and $J_h$ (blue hexagons).
            In this work, dashed (solid) bonds correspond to ferromagnetic (antiferromagnetic) interactions.
        }
\end{figure}
Different inter-atomic distances and exchange mechanisms in crystals generically create both ferromagnetic (FM) and antiferromagnetic (AFM) interactions between local spin degrees of freedom.
A common result of this interplay are frustrated ferromagnets whose phase transitions and instabilities are often highly nontrivial.
A recurring theme in this context is that direct FM~$\leftrightarrow$~AFM transitions occurring in a classical model can be accompanied by intermediate, e.g. spin-nematic or lattice-nematic, phases in the quantum case~\cite{starykh:2004, Shannon2006, Iqbal-2016,   richter:2010, Starykh2015, Jiang2023QSN}.
On the other hand, there are scenarios in which a classically existing intermediate phase vanishes in the presence of quantum many-body effects, leading either to a direct transition~\cite{jiang:2023b} or giving rise to one or more new putative quantum phases~\cite{lauchli:2009, jiang:2023b}.
In extreme cases of frustration, the mechanisms by which the ferromagnetic phase becomes unstable can be rather exotic.
For instance, octupolar order was shown to arise on the triangular lattice due to three-magnon interactions~\cite{momoi:2006} and a flat one-magnon band is known to exist above the FM state on the kagome lattice leading to so-called ``magnon crystals''~\cite{Derzhko:2015, zhitomirsky:2004}.

The maple-leaf lattice (MLL)~\cite{betts:1995}, which is illustrated in Fig.~\ref{fig:lattice}, is a site-depleted triangular lattice with coordination number five and no spatial reflection symmetry.
\begin{figure*}[t]
        \centering
        \includegraphics[width=\linewidth]{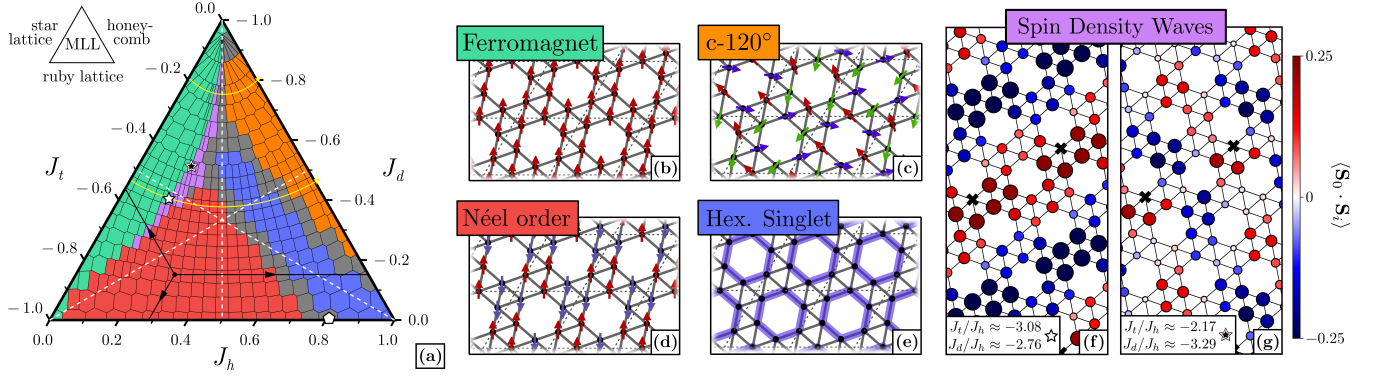}
        \caption{\label{fig:phase_diagram}
            (a): Color-coded ternary ($|J_t| + J_h + |J_d| = 1$) phase diagram of the Hamiltonian~\eqref{eq:H} on the MLL based on $N=36$ ED data (shown as Voronoi tessellation), where the couplings $J_t, J_d \leq 0$ are ferromagnetic and $0 \leq J_h$ is antiferromagnetic.
            Data points which could not be conclusively attributed to a phase are drawn in gray.
            The three black lines with arrows pointing away from a common point illustrate how couplings should be read off.
            The paths of the phase diagram cuts shown in Fig.~\ref{fig:N-36-N1-N4-cut} are drawn as yellow arrows.
            As illustrated to the left of the diagram, the three external boundaries correspond to Heisenberg models on the ruby, star, and honeycomb lattices, respectively obtained by setting $J_d$, $J_h$ or $J_t$ to zero.
            Lines along which the magnitude of two couplings is equal are drawn in white, their intersection marking the ``isotropic'' point where $J_t = J_d = - J_h < 0$.
            The white pentagon in the bottom right of the phase diagram highlights the point at which the ED ground state data agrees well with the Gutzwiller-projected quantum spin liquid Ansatz Z1012 as classified in Ref.~\cite{Sonnenschein2024}. 
            (b)-(e): Illustrations of the identified ferromagnetic, \canted{}, collinear N\'eel, and hexagonal singlet phases.
            (f)-(g): Spin-density-wave regime exemplified by $N=36$ spin-spin correlator data at the two points highlighted by stars inside the diagram (a). 
        }
\end{figure*}
It has recently gained significant theoretical and experimental attention in the context of exhibiting an exact ($J_d$) dimer eigenstate~\cite{Ghosh2022}, coplanar orders~\cite{farnell:2011, farnell:2014, beck:2024, gresista:2023, Gresista:2026, gembe:2024, richter:2010, schmoll:2025, Ebert2026, ghosh:2025b, Aguilar-Maldonado:2025, Haraguchi2018, saha:2023, nakano:2026}, magnetization plateaus~\cite{Ghosh2023Field, ghosh:2024c, schmoll:2024a, Haraguchi2018}, anomalous thermodynamic signatures~\cite{venkatesh:2020, haraguchi:2021, Schaefer:2026, saha:2023, borutta:2026} as well as potential quantum spin liquid (QSL) phases~\cite{gresista:2023, beck:2024, schmoll:2025, gembe:2024, Ebert2026, Sonnenschein2024, ghosh:2024a}.
Indeed, several known MLL materials such as Bluebellite~\cite{haraguchi:2021, mills:2014, ghosh:2024c} or Spangolite~\cite{fennell:2011, schmoll:2024a} approximately realize $S=1/2$ ferro-antiferromagnetic nearest-neighbor Heisenberg models such as
\begin{equation}\label{eq:H}
        H = 
        J_t \sum_{\langle i, j\rangle \in t}   \bm{S}_i \cdot \bm{S}_j 
        + J_h \sum_{\langle i, j\rangle \in h}   \bm{S}_i \cdot \bm{S}_j
        + J_d \sum_{\langle i, j\rangle \in d}   \bm{S}_i \cdot \bm{S}_j,
\end{equation}
with three (or more) independent nearest-neighbor couplings $J_t, J_h, J_d$ as defined in Fig.~\ref{fig:lattice}.
Spangolite is known to essentially host a $J_d$-dimer ground state that is dressed by ferromagnetic $J_h$ and antiferromagnetic $J_t$ interactions~\cite{schmoll:2024a}.
For Bluebellite it was argued that the strong ferro-antiferromagnetic interactions alternating around hexagons ($J_{h1} = - J_{h2}$) lead to an effective $S=1$ kagome model with valence-bond-solid ground state~\cite{ghosh:2024c}.
Recently Na$_2$Mn$_3$O$_7$, realizing a $S=3/2$ MLL system, was investigated experimentally~\cite{venkatesh:2020, saha:2023} and theoretically~\cite{borutta:2026} with reports converging on strongly antiferromagnetic $0 \ll J_h$ and otherwise mostly ferromagnetic $J_t, J_d < 0$ interactions.
The $J_t, J_d < 0 < J_h$ phase diagram of the Hamiltonian~\eqref{eq:H} is thus the natural test-bed for studying frustrated maple-leaf ferromagnets.

In this Letter, we discuss the $J_t, J_d \leq 0 \leq J_h$ phase diagram of the $S=1/2$ Heisenberg model~\eqref{eq:H}.
We show that it exhibits FM, collinear N\'eel AFM, hexagonal singlet (HS), and canted $120^{\circ}$ (\canted{}) phases, as well as a large spin-density-wave regime close to the FM boundary in which the ordering vector varies continuously and energy levels are compressed by orders of magnitude.
This is to be contrasted to the previously hypothesized zero-field nematic phase~\cite{Gresista:2026}, for which we do not find direct finite-size evidence.
Furthermore, our data does not support the possibility of a dimerized hexagonal-singlet ground state~\cite{Gresista:2026, ghosh:2024a}.

\paragraph*{Methods.}
    Our analysis is based on exact diagonalization (ED) using the \texttt{XDiag} library~\cite{wietek:2025:xdiag:paper, wietek:2025:xdiag:code} on size $N\in\{42, 36, 24, 18\}$ clusters.
    We resolve all irreducible representations (irreps) of their respective space groups as well as different values of the total $S^z_{\mathrm{tot}} = \sum_i S_i^z$ magnetization, giving access to the SU(2) quantum number $S$.
    Our identification of phases is based on the energy level and irrep structure of ED spectra, various ground state correlators, in particular structure factors, as well as tower-of-states (TOS) analysis.
    Furthermore, we perform targeted variational Monte Carlo studies of the candidate spin-liquid states classified in Ref.~\cite{Sonnenschein2024} on the same set of clusters, the latter being discussed in Sec. C of the Supplemental Material (SM)~\cite{SupplementalMaterial}.

\paragraph*{Phase Diagram.}
    
    The resulting phase diagram of the $S=1/2$ Hamiltonian~\eqref{eq:H} with $J_t, J_d \leq 0 \leq J_h$ is shown in the ternary plot of Fig.~\ref{fig:phase_diagram}{a}. 
    In agreement with Ref.~\cite{Gresista:2026} and the corner-sharing $0 \leq J_t, J_h, J_d$ phase diagram~\cite{Ebert2026}, we find a FM, a collinear N\'eel AFM, a  \canted{} order as well as the HS phase.
    At some of the scanned points, highlighted in gray in Fig.~\ref{fig:phase_diagram}, we were unable to unambiguously classify the nature of the realized phase.
    For most of the collinear N\'eel $\leftrightarrow$ HS or the HS $\leftrightarrow$ \canted{} phase boundaries we do not expect intermediate phases although these regions are shown in gray.
    In particular, the dimerized hexagonal-singlet (d-HS) state (three dimers on each $J_h$-hexagon, breaking the point group from $C_6$ to $C_3$), conjectured to exist close to the boundary of the \canted{} phases in Ref.~\cite{Gresista:2026} is not observed on $N \in \{36, 24, 18\}$ clusters.
    This not only follows from ground state correlators, but also from the fact that point-group-violating dimer states would be expected to produce a set of low-lying singlet states associated with the broken point-group symmetry, which are absent.
    As discussed in detail below, we see no indications for a spin-nematic phase near the FM boundary, as put forward in Ref.~\cite{Gresista:2026}, but see clear signs of a spin-density wave instability in the purple region of Fig.~\ref{fig:phase_diagram}{a}.

    \begin{figure}
        \centering
        \includegraphics[width=\linewidth]{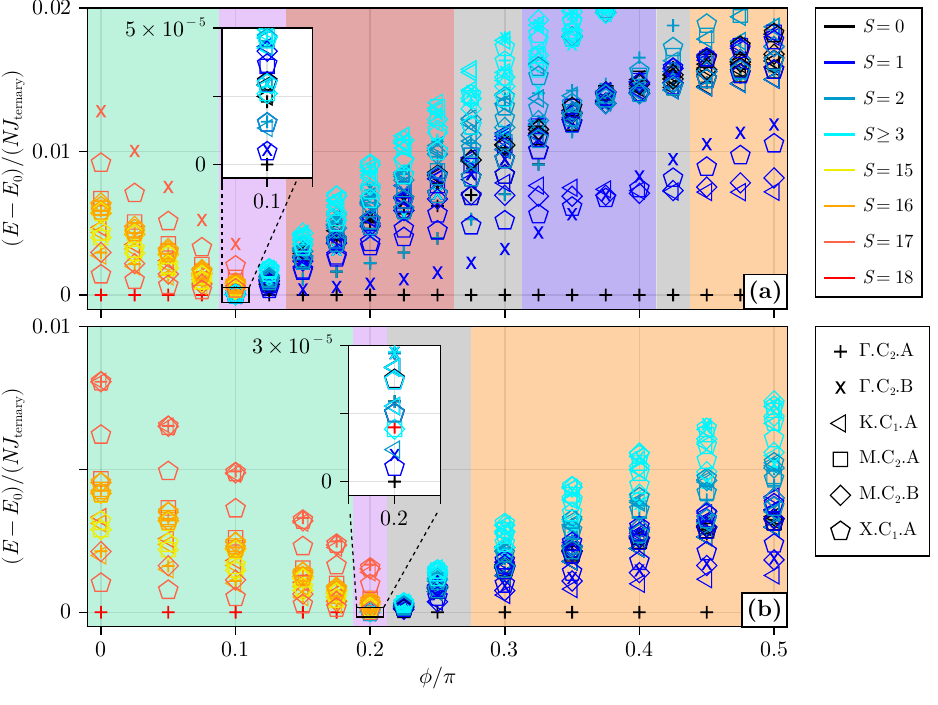}
        \caption{\label{fig:N-36-N1-N4-cut}
            Lowest energy levels along the two cuts illustrated in Fig.~\ref{fig:phase_diagram}{a} as yellow arrows; (a) lower arrow at intermediate $J_d$; (b) upper arrow at strong $J_d$.
            The angle $\phi$ originates from a spherical parametrization of the phase diagram in Fig.~\ref{fig:phase_diagram}{a}; see Sec. D in the SM~\cite{SupplementalMaterial}.
            The energy scale is chosen such that $1 = J_{\rm ternary} = |J_t| + J_h + |J_d|$.
            Background colors highlight the classification of phases in the color scheme of Fig.~\ref{fig:phase_diagram}.
            Space-group irreps are fully resolved and $\text{SU}(2)$ irreps labelled by $S$ are shown for $S \leq 3$ outside and $S \geq 15$ inside the FM phase.
            Note that we cannot fully distinguish between $S = 3$ and $3 < S < 15$ levels without diagonalizing all $S^z_{\rm tot}=4$ representations, i.e., some of the $S\geq 3$ levels may actually belong to $3 < S < 15$ sectors.
            Note the strong compression of energy levels in the spin-density-wave region and the fact that an $S=18$ level becomes one of the first excited states in the (b) inset (full $E_0(S)$ spectra in Fig.~\ref{fig:app:tos} in the SM~\cite{SupplementalMaterial}).
        }
    \end{figure}

     \begin{figure}
        \centering
        \includegraphics[width=\linewidth]{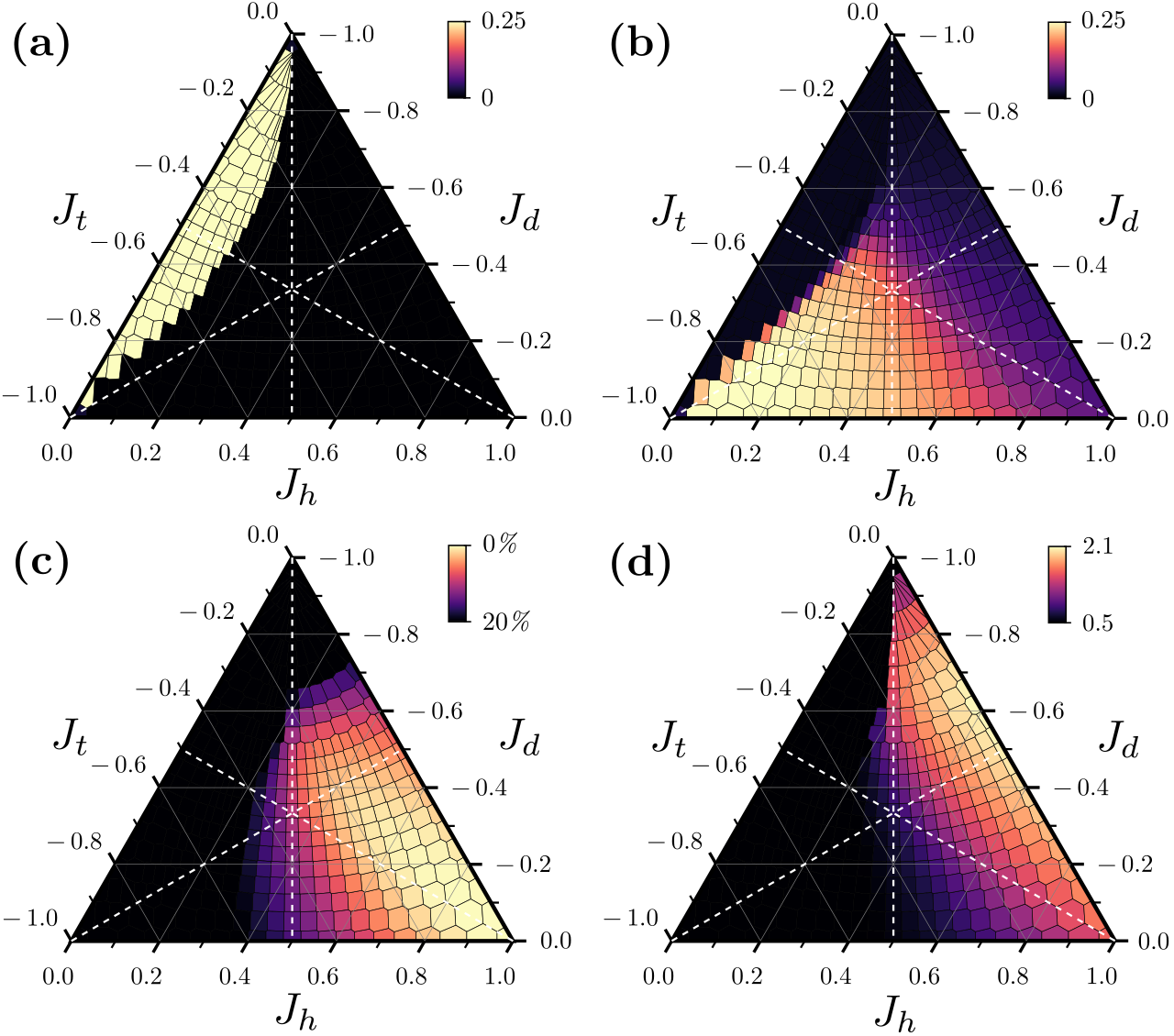}
        \caption{\label{fig:N-36-correlators}
            Ground state correlators on the $N=36$ MLL cluster.
            (a): Spin-spin correlator from Eq.~\eqref{eq:FM_op};
            (b): Staggered spin-spin correlator from Eq.~\eqref{eq:AFM_op};
            (c): ``Hexagonal singlet correlator'' $\Delta_{\text{HS}}$ from Eq.~\eqref{eq:HS_op};
            (d): Structure factor $\mathcal{S}(K_{\text{TRI}})$.
        }
    \end{figure}
    
    To illustrate how different phases can be inferred from the structure of low-lying eigenstates, two horizontal cuts, highlighted in yellow in Fig.~\ref{fig:phase_diagram}{a}, are shown in Fig.~\ref{fig:N-36-N1-N4-cut}.
    As expected, the ground state in the \fm{} phase belongs to the fully polarized $S = S_\text{max} = N/2$ sector with the first and second excited states being $S_\text{max} -1$ and $S_\text{max} - 2$ levels respectively.
    In all other cases, an $S=0$ (singlet) level is found to be the ground state and the excited states contain information about the realized phase.
    For instance, the collinear N\'eel (red background) region characteristically features a $\Gamma$.C$_2$.B triplet ($S=1$) and a $\Gamma$.C$_2$.A quintuplet ($S=2$), which matches the analytical TOS prediction (see Tab.~1 in Ref.~\cite{Ebert2026}).
    The hexagonal singlet is characterized by a clear separation of energy levels according to $S$ as well as a low-lying $\Gamma$.C$_2$.B triplet level.
    Lastly, the TOS of the \canted{} order can be analytically shown (see Ref.~\cite{Ebert2026}) to be comprised of $\Gamma$.C$_6$.A, $\Gamma$.C$_6$.B, and K.C$_3$.A levels in the thermodynamic limit, i.e., for each $S$ on the $N=36$ cluster we expect the lowest $S$ level to be part of the $\Gamma$.C$_2$.A, $\Gamma$.C$_2$.B, or K.C$_1$.A irrep (as the finite cluster does not resolve the full little co-groups of each high symmetry point).

\begin{figure}[t]
        \centering
        \includegraphics[width=\linewidth]{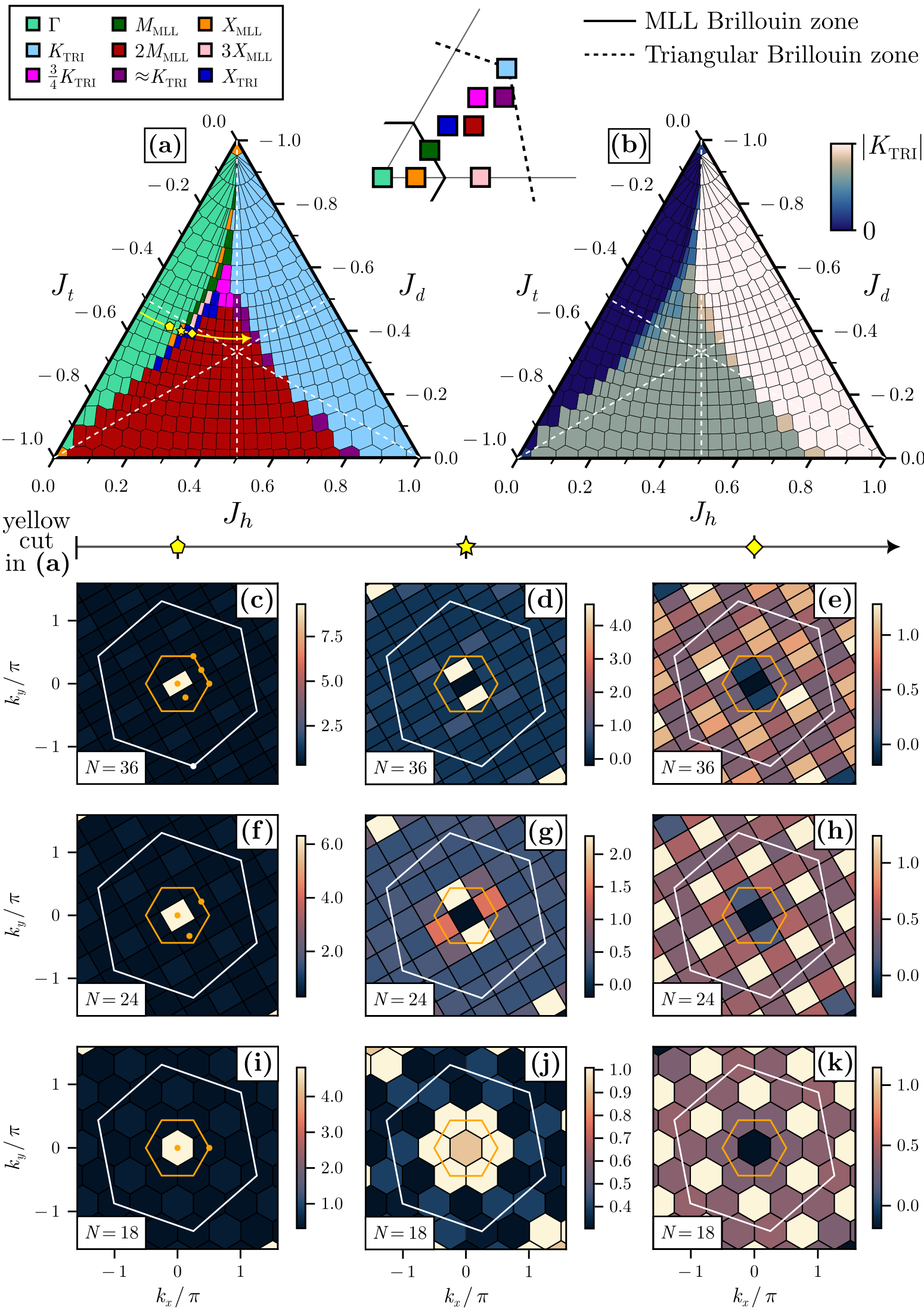}
        \caption{\label{fig:structureFactor}
            (a): Map of reciprocal-lattice vectors $\bm{q}_{\mathrm{max}}$ at which the structure factor $\mathcal{S}(\bm{q})$ peaks on the $N=36$ cluster, with an illustration of the extended Brillouin zone.
            (b): Distance from the $\mathcal{S}(\bm{q})$-maximizing reciprocal-lattice vector $\bm{q}_{\mathrm{max}}$ on the $N=36$ cluster to the $\Gamma$ point, i.e., $|\bm{q}_{\mathrm{max}}|$;
            (c)-(k): Structure factors along the yellow cut in (a) are shown in (c)-(e) for $N=36$,  (f)-(h) for $N=24$,  (i)-(k) for $N=18$.
        }
    \end{figure}

    Further evidence from ground state correlators is shown in Fig.~\ref{fig:N-36-correlators}, respectively displaying the uniform (a) and staggered magnetization (b):
    \begin{align}
        m &= \frac{1}{N} \sum_{i = 0}^{N-1} \langle \bm{S}_0 \cdot \bm{S}_i\rangle,\label{eq:FM_op}\\
        m_s &= \frac{1}{N} \sum_{i = 0}^{N-1} (-1)^i \langle \bm{S}_0  \cdot \bm{S}_i\rangle,\label{eq:AFM_op}
    \end{align}
    where $(-1)^i$ is used to denote the bipartition of the $J_t = 0$ (honeycomb) graph (see~ Fig.~\ref{fig:phase_diagram}{d}).
    Here spin-spin correlators are needed since single-spin measurements are constrained by the fixed $S^z_{\rm tot}$ value.
    We find that the cleanest way of diagnosing the HS phase through ground state correlators is to compare the ED spin-spin correlations on $J_h$ bonds to the value $e_{\text{HS}}$ expected on isolated six-spin $J_h$ rings
    \begin{equation}\label{eq:HS_op}
        \Delta_{\text{HS}} = \frac{1}{N} \sum_{\langle i, j \rangle\in h} \frac{|\langle \bm{S}_i\cdot \bm{S}_j\rangle -e_{\text{HS}}|}{|e_{\text{HS}}|},
    \end{equation}
    which is shown in (c) of Fig.~\ref{fig:N-36-correlators} and is near zero in the HS phase by construction.
    Although $\Delta_{\text{HS}}$ starts to deviate notably ($\approx 10 \%$) from the pure HS spin-spin correlations north of the ``isotropic point'' we still attribute some of these data points to the HS phase as the ED spectra are still compatible with this scenario and spin-spin correlations are still dominated by antiferromagnetic intra-hexagon contributions and are near zero everywhere else.
    Lastly, it is well established through various numerical approaches~\cite{Gresista:2026, beck:2024, gembe:2024, Schaefer:2026, gresista:2023, Ebert2026} that the \canted{} order features strong peaks of the equal-time spin-spins structure factor $\mathcal{S}(\bm{q})$ at the high-symmetry momentum $\text{K}_{\text{TRI}}$ of the Brillouin zone of the underlying triangular lattice~\footnote{
        Recall that the maple-leaf lattice can be obtained by a one-seventh site depletion of the triangular lattice.
    } which lies inside the extended Brillouin zone of the MLL (see Fig.~\ref{fig:structureFactor}).
    We thus show $\mathcal{S}(\text{K}_{\text{TRI}})$ in subplot (d), where the observed peak values are lower than in the purely antiferromagnetic Heisenberg model~\cite{Ebert2026}, however.
    This may hint at a stronger canting away from the uniform $120^\circ$ order in the presence of FM interactions, which is supported by a very weak response of the ``twist-twist'' correlator~\cite{Ebert2026} in that region, otherwise signaling a uniform $120^{\circ}$ degree order.

\paragraph*{Instabilities of the Zero-Field Ferromagnet}

    The recent pf-FRG study~\cite{Gresista:2026} identified a region close to the FM boundary where the structure factor $\mathcal{S}(\bm{q})$ peaks along a continuous ring inside the MLL Brillouin zone.
    Combined with a strong response under an explicit breaking of the SU(2) spin symmetry down to U(1), this led to the hypothesis of a spin-nematic phase near the FM boundary~\cite{Gresista:2026}.
    It is well-established that quadrupolar (octupolar) spin-nematic phases are related to two-magnon (three-magnon) interactions and can hence be identified through low-lying $S=2$ ($S=3$) levels just outside the FM phase~\cite{wietek:2020, penc:2011, Shannon2006, momoi:2006}.
    Despite extensive high-resolution ED scans of the FM boundary on $N\in \{36, 24, 18\}$ clusters, we could not identify a single point at which such behavior is manifest.
    Moreover, on the $N=36$ system we find the two-magnon interaction to be near zero or repulsive throughout the purple region in Fig.~\ref{fig:phase_diagram}, and thus no evidence for a zero-field $d$-wave spin-nematic phase on the clusters studied here~\cite{Shannon2006, wietek:2020} (data shown in Sec. A of the SM~\cite{SupplementalMaterial}).
    Instead, along the whole FM boundary we find ED spectra as in Fig.~\ref{fig:N-36-N1-N4-cut}, where the low-energy spectrum collapses into a narrow window, with excitation energies reduced by more than an order of magnitude compared to neighboring phases, as the ground state flips from the $S = S_\mathrm{max} = N/2$ to the $S=0$ sector.
    Right after the transition, a regime of stripe orders is found where $\mathcal{S}(\bm{q})$ peaks at different momenta $\bm{q}_\mathrm{max}$ between the $\Gamma$ point (FM phase) and the ordering vector of the respective proximate phase ($2 M_{\mathrm{MLL}}$ for the collinear N\'eel AFM and $K_\mathrm{TRI}$ for the \canted{} order).
    This continuous movement of the ordering vector away from $\Gamma$ simultaneously occurs on clusters of different sizes as demonstrated in Fig.~\ref{fig:structureFactor} along the cut shown in Fig.~\ref{fig:N-36-N1-N4-cut}{a} for $N=36$ in (c)-(e), for $N=24$ in (f)-(h), and for $N=18$ in (i)-(k).
    On the available clusters, this intermediate stripe regime is at least as prominent on the larger cluster as on the smaller ones, as exemplified by the fact that the structure factor in (e) has not yet reached the $2M_{\rm MLL}$ point of the collinear N\'eel AFM phase, while (h) and (k) already show these characteristic N\'eel peaks.
    Subfigure (b) in Fig.~\ref{fig:structureFactor} further shows the continuous nature of the structure factor peak moving away from $\Gamma$. 
    Spin-spin correlations of two representative stripe patterns are shown in (f) and (g) of Fig.~\ref{fig:phase_diagram}.
    The same regions are characterized by an unusually small TOS slope, indicating a large susceptibility to magnetic fields (see Fig.~\ref{fig:app:tos}{a} in the SM~\cite{SupplementalMaterial}).
    At one point in the $J_t / J_h \lesssim 0, J_d / J_h \ll 0$ regime particularly close to the FM phase we even observe an arch-shaped behavior of the ground state energy $E_0(S)$ inside a fixed-$S$ sector (see Fig.~\ref{fig:app:tos}{b} in the SM~\cite{SupplementalMaterial}).
    If $E_0(S)$ retains this arch-shape with increasing system size, it signals a first-order transition at the upper regions of the FM boundary in Fig.~\ref{fig:phase_diagram} since the ground state then has to jump from the $S\approx 0$ sector into $S=S_{\rm max}$.
    Thus, while the spectra show strong softening near the FM boundary, the finite-size ordering of spin sectors is closer to a soft incommensurate instability than to the standard bound-magnon route to a zero-field quadrupolar phase.
  
    \begin{figure}
        \centering
        \includegraphics[width=\linewidth]{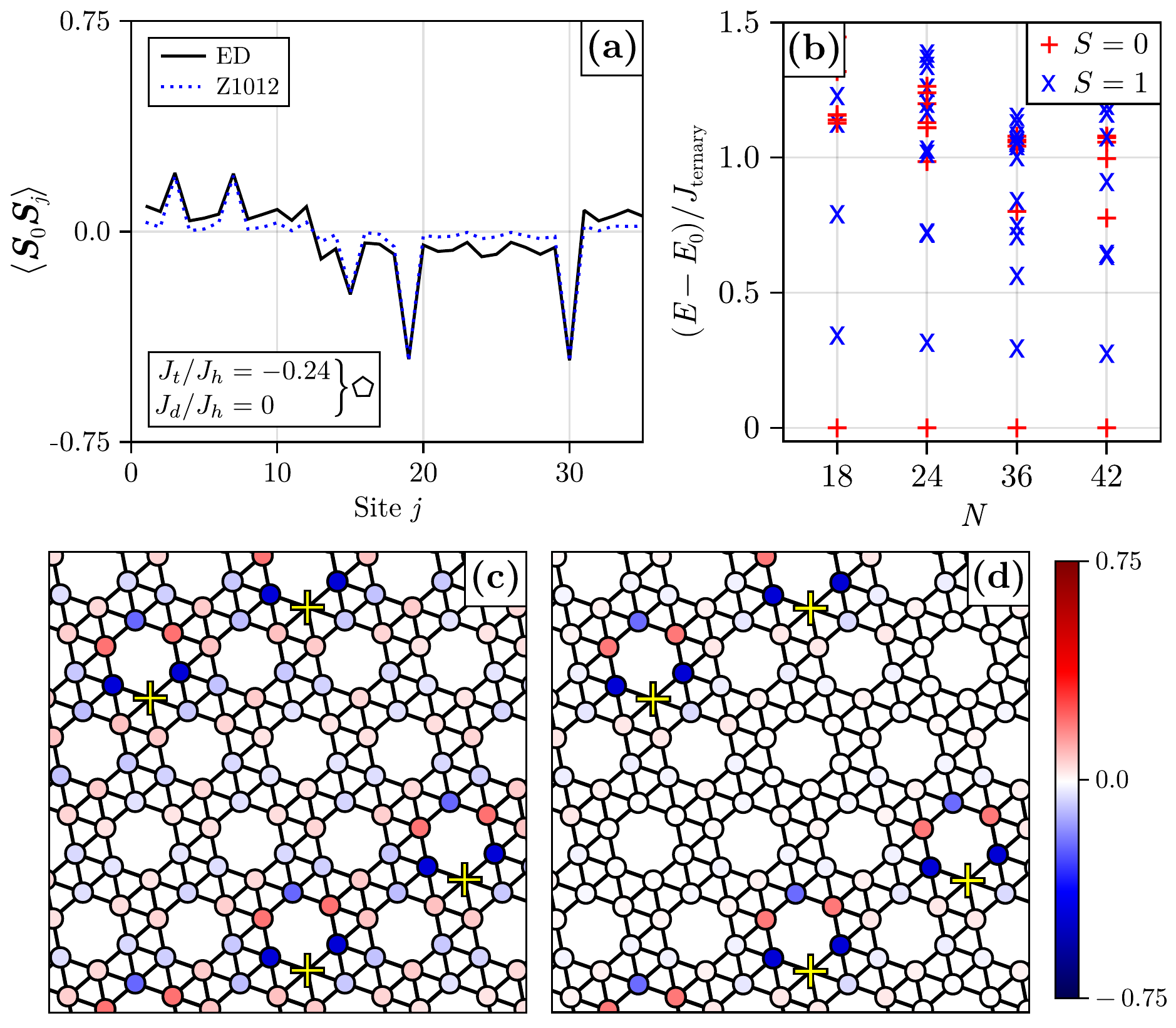}
        \caption{\label{fig:QSL} (a) Spin-spin correlations in the ED ground state compared to the Gutzwiller-projected $\mathbb{Z}_2$ QSL Ansatz Z1012 on $N=36$ sites at the point marked by a white pentagon in the phase diagram in Fig.~\ref{fig:phase_diagram}.
        (b): Low-lying energy levels at the same point as in (a) but on different system sizes. The energy scale is chosen such that $1 = J_{\rm ternary} = |J_t| + J_h + |J_d|$.
        (c) and (d): Spin-spin correlations from (a) drawn in real space with periodic boundary conditions for ED (c) and the Z1012 Ansatz (d).
        }
    \end{figure}

\paragraph*{Comparison to Fermionic Spin-Liquid {\it Ans\"atze}.}

    Ref.~\cite{Sonnenschein2024} classified all fully symmetric Abrikosov-fermion U$(1)$ and $\mathbb{Z}_2$ QSL {\it Ans\"atze} on the MLL. We test these states by variational Monte Carlo (VMC), using Gutzwiller-projected wave functions to compare their energies and spin-spin correlations to the ED ground state.
    It is known that these variational descriptions work well for the U(1) Dirac spin-liquids on the kagome and triangular lattices~\cite{Iqbal-2013,Iqbal-2016, wietek:2024}, reproducing the ED ground state with 
    \begin{equation}
        \Delta_\mathrm{VMC} = |(E_{\rm VMC}-E_{\rm ED})/E_{\rm ED}|
    \end{equation}
    of a few percent.
    Recently, we also reported on a promising regime in the nearest-neighbor maple-leaf fully antiferromagnetic model~\cite{Ebert2026} where two Gutzwiller-projected $\mathbb{Z}_2$ {\it Ans\"atze} reach $\Delta_\mathrm{VMC}\sim 10^{-3}$.

    We repeat this analysis in the present ferro-antiferromagnetic setting, studying all twelve $\mathrm{U}(1)$ and eight $\mathbb{Z}_{2}$ wave functions~\cite{Sonnenschein2024} at representative coupling values in the gray regions between the classified phases in Fig.~\ref{fig:phase_diagram}{a}.
At none of the interior points considered do these VMC wave functions reach competitive energies on the same $N=36$ cluster.
This distinguishes the gray unresolved regions from the fully symmetric fermionic spin-liquid manifold already classified for this lattice~\cite{Sonnenschein2024}, rather than leaving it uncharacterized.

An exception is found on the $J_d=0$ ruby (bounce) lattice boundary, at an intermediate point between the collinear N\'eel and HS phases marked by the white pentagon in Fig.~\ref{fig:phase_diagram}{a}.
At this point, the $\mathbb{Z}_2$ Ansatz Z1012, in the nomenclature of Ref.~\cite{Sonnenschein2024}, yields a gapped wave function with an energy mismatch below $1\%$, i.e., $\Delta_{\rm VMC}<10^{-2}$.
Moreover, the spin-spin correlations of the ED ground state are qualitatively well reproduced, as shown in Fig.~\ref{fig:QSL}, although the quantitative agreement is less precise than in the previously studied purely antiferromagnetic model~\cite{Ebert2026}.
The ED spectra at this point are displayed in Fig.~\ref{fig:QSL}{b} across different system sizes, showing that non-trivial low-lying singlet ($S=0$) levels are present between the collinear N\'eel and HS phases on all clusters up to $N=42$.
It is thus possible that more singlet levels appear on larger clusters or that an extended model realizes a nearby QSL phase.

\paragraph*{Discussion and Outlook.}

We investigated the phase diagram of the $J_t, J_d \leq 0 \leq J_h$ nearest-neighbor Heisenberg model on the maple-leaf lattice using symmetry-resolved exact diagonalization and tower-of-states analysis.
While the ferromagnetic, collinear N\'eel antiferromagnetic, canted $120^{\circ}$, and hexagonal singlet phases were identified in agreement with another recent study~\cite{Gresista:2026}, our data on $N \in \{36,24,18\}$ clusters do not support direct finite-size evidence for either of the hypothesized scenarios of a dimerized hexagonal singlet or a zero-field spin-nematic phase.
The former is a dimer phase breaking the point group and must thus feature a low-lying singlet ($S=0$) state, which is not observed in the respective regions of the phase diagram.
On the other hand, quadrupolar (octupolar) spin-nematic behavior is expected to be accompanied by low-lying $S=2$ ($S=3$) levels due to two-magnon (three-magnon) interactions~\cite{momoi:2006, Shannon2006, wietek:2020}, which are not observed anywhere along the boundary of the ferromagnetic phase and the two-magnon interaction appears to be repulsive.
Instead, our data suggest that almost the entirety of the ferromagnetic phase in the $J_t, J_d \leq 0 \leq J_h$ Heisenberg model is destabilized by spin density waves whose ordering vectors continuously shift away from $\Gamma$ to that of the neighboring magnetically ordered phases.

The high-$S$ states of exact diagonalization spectra near the ferromagnetic boundary further display no anomalous behavior (such as an odd-even periodicity in the tower of states~\cite{Shannon2006}; data shown in Fig.~\ref{fig:app:tos}) which renders a spin-nematic phase induced through magnetic fields near saturation unlikely.
Overall, the degree of frustration in the present $J_t, J_d \leq 0 \leq J_h$ model appears to be too weak to introduce the same type of multipolar or flat-band physics found on the triangular or kagome lattices~\cite{momoi:2006, Derzhko:2015, zhitomirsky:2004}.
The naturally emerging question is whether this is due to having only one antiferromagnetic coupling or whether the maple-leaf lattice itself lacks the sufficient degree of frustration.

Regarding materials, the presented $S=1/2$ phase diagram is consistent with recent results on Na$_2$Mn$_3$O$_7$ which realizes a $S=3/2$ maple-leaf geometry~\cite{borutta:2026, saha:2023, venkatesh:2020}.
If the nine nearest-neighbor interactions obtained in Ref.~\cite{borutta:2026} are averaged into three couplings values, one obtains (under ternary normalization) $J_h \approx 0.8$, $J_d \approx -0.12$, $J_t \approx -0.07$.
Na$_2$Mn$_3$O$_7$ would thus be in the bottom right corner of Fig.~\ref{fig:phase_diagram}{a} and inside the hexagonal singlet phase if it were a $S=1/2$ material instead of $S=3/2$.
Our data thus match Ref.~\cite{borutta:2026} reporting nearly isolated antiferromagnetic hexagons and rapidly decaying inter-hexagon correlations in the $S=3/2$ case.

Using variational Monte Carlo we obtained Gutzwiller-projected wave functions of the fully symmetric fermionic quantum spin liquid Ans\"atze classified in Ref.~\cite{Sonnenschein2024}.
We identify a point in the $J_d = 0$ ruby lattice limit at $J_t / J_h \approx -0.24$ between the collinear N\'eel and hexagonal singlet phases that is well described by a gapped $\mathbb{Z}_2$ Ansatz.
This result could hint at a $\mathbb{Z}_2$ quantum spin liquid being realized at a nearby point in an extension of ferro-antiferromagnetic nearest-neighbor Heisenberg model on the ruby lattice -- a geometry already realized in contemporary programmable Rydberg-array quantum-simulator platforms~\cite{semeghini:2021} and for which non-Heisenberg $\mathbb{Z}_2$ QSL phases are well-explored~\cite{giudici:2022, samajdar:2023, tarabunga:2022}.

\paragraph*{Acknowledgments.}
We thank Karlo Penc for very helpful discussions. A.W. acknowledges support by the German Research Foundation (DFG) through the Emmy Noether program (Grant No. 509755282), IIT Madras for a Visiting Faculty Fellow position under the IoE program, and the European
Research Council (ERC) under the European Union’s Horizon
Europe research and innovation program (Project ID 101220368)—ERC Starting Grant MoNiKa. The work of Y.I. was performed in part at the Aspen Center for Physics, which is supported by a grant from the Simons Foundation (1161654, Troyer). This research was also supported in part by grant NSF PHY-2309135 to the Kavli Institute for Theoretical Physics and by the International Centre for Theoretical Sciences through participation in the Discussion Meeting --- Fractionalized Quantum Matter (code: ICTS/DMFQM2025/07). Y.I. also acknowledges the use of the computing resources at HPCE, IIT Madras.

\bibliography{mll}

\begin{thebibliography}{49}%
\makeatletter
\providecommand \@ifxundefined [1]{%
 \@ifx{#1\undefined}
}%
\providecommand \@ifnum [1]{%
 \ifnum #1\expandafter \@firstoftwo
 \else \expandafter \@secondoftwo
 \fi
}%
\providecommand \@ifx [1]{%
 \ifx #1\expandafter \@firstoftwo
 \else \expandafter \@secondoftwo
 \fi
}%
\providecommand \natexlab [1]{#1}%
\providecommand \enquote  [1]{``#1''}%
\providecommand \bibnamefont  [1]{#1}%
\providecommand \bibfnamefont [1]{#1}%
\providecommand \citenamefont [1]{#1}%
\providecommand \href@noop [0]{\@secondoftwo}%
\providecommand \href [0]{\begingroup \@sanitize@url \@href}%
\providecommand \@href[1]{\@@startlink{#1}\@@href}%
\providecommand \@@href[1]{\endgroup#1\@@endlink}%
\providecommand \@sanitize@url [0]{\catcode `\\12\catcode `\$12\catcode `\&12\catcode `\#12\catcode `\^12\catcode `\_12\catcode `\%12\relax}%
\providecommand \@@startlink[1]{}%
\providecommand \@@endlink[0]{}%
\providecommand \url  [0]{\begingroup\@sanitize@url \@url }%
\providecommand \@url [1]{\endgroup\@href {#1}{\urlprefix }}%
\providecommand \urlprefix  [0]{URL }%
\providecommand \Eprint [0]{\href }%
\providecommand \doibase [0]{https://doi.org/}%
\providecommand \selectlanguage [0]{\@gobble}%
\providecommand \bibinfo  [0]{\@secondoftwo}%
\providecommand \bibfield  [0]{\@secondoftwo}%
\providecommand \translation [1]{[#1]}%
\providecommand \BibitemOpen [0]{}%
\providecommand \bibitemStop [0]{}%
\providecommand \bibitemNoStop [0]{.\EOS\space}%
\providecommand \EOS [0]{\spacefactor3000\relax}%
\providecommand \BibitemShut  [1]{\csname bibitem#1\endcsname}%
\let\auto@bib@innerbib\@empty
\bibitem [{\citenamefont {Starykh}\ and\ \citenamefont {Balents}(2004)}]{starykh:2004}%
  \BibitemOpen
  \bibfield  {author} {\bibinfo {author} {\bibfnamefont {O.~A.}\ \bibnamefont {Starykh}}\ and\ \bibinfo {author} {\bibfnamefont {L.}~\bibnamefont {Balents}},\ }\bibfield  {title} {\bibinfo {title} {Dimerized phase and transitions in a spatially anisotropic square lattice antiferromagnet},\ }\href {https://doi.org/10.1103/PhysRevLett.93.127202} {\bibfield  {journal} {\bibinfo  {journal} {Phys. Rev. Lett.}\ }\textbf {\bibinfo {volume} {93}},\ \bibinfo {pages} {127202} (\bibinfo {year} {2004})}\BibitemShut {NoStop}%
\bibitem [{\citenamefont {Shannon}\ \emph {et~al.}(2006)\citenamefont {Shannon}, \citenamefont {Momoi},\ and\ \citenamefont {Sindzingre}}]{Shannon2006}%
  \BibitemOpen
  \bibfield  {author} {\bibinfo {author} {\bibfnamefont {N.}~\bibnamefont {Shannon}}, \bibinfo {author} {\bibfnamefont {T.}~\bibnamefont {Momoi}},\ and\ \bibinfo {author} {\bibfnamefont {P.}~\bibnamefont {Sindzingre}},\ }\bibfield  {title} {\bibinfo {title} {{Nematic Order in Square Lattice Frustrated Ferromagnets}},\ }\href {https://doi.org/10.1103/PhysRevLett.96.027213} {\bibfield  {journal} {\bibinfo  {journal} {Phys. Rev. Lett.}\ }\textbf {\bibinfo {volume} {96}},\ \bibinfo {pages} {027213} (\bibinfo {year} {2006})}\BibitemShut {NoStop}%
\bibitem [{\citenamefont {Iqbal}\ \emph {et~al.}(2016)\citenamefont {Iqbal}, \citenamefont {Ghosh}, \citenamefont {Narayanan}, \citenamefont {Kumar}, \citenamefont {Reuther},\ and\ \citenamefont {Thomale}}]{Iqbal-2016}%
  \BibitemOpen
  \bibfield  {author} {\bibinfo {author} {\bibfnamefont {Y.}~\bibnamefont {Iqbal}}, \bibinfo {author} {\bibfnamefont {P.}~\bibnamefont {Ghosh}}, \bibinfo {author} {\bibfnamefont {R.}~\bibnamefont {Narayanan}}, \bibinfo {author} {\bibfnamefont {B.}~\bibnamefont {Kumar}}, \bibinfo {author} {\bibfnamefont {J.}~\bibnamefont {Reuther}},\ and\ \bibinfo {author} {\bibfnamefont {R.}~\bibnamefont {Thomale}},\ }\bibfield  {title} {\bibinfo {title} {{Intertwined nematic orders in a frustrated ferromagnet}},\ }\href {https://doi.org/10.1103/PhysRevB.94.224403} {\bibfield  {journal} {\bibinfo  {journal} {Phys. Rev. B}\ }\textbf {\bibinfo {volume} {94}},\ \bibinfo {pages} {224403} (\bibinfo {year} {2016})}\BibitemShut {NoStop}%
\bibitem [{\citenamefont {Richter}\ \emph {et~al.}(2010)\citenamefont {Richter}, \citenamefont {Darradi}, \citenamefont {Schulenburg}, \citenamefont {Farnell},\ and\ \citenamefont {Rosner}}]{richter:2010}%
  \BibitemOpen
  \bibfield  {author} {\bibinfo {author} {\bibfnamefont {J.}~\bibnamefont {Richter}}, \bibinfo {author} {\bibfnamefont {R.}~\bibnamefont {Darradi}}, \bibinfo {author} {\bibfnamefont {J.}~\bibnamefont {Schulenburg}}, \bibinfo {author} {\bibfnamefont {D.~J.~J.}\ \bibnamefont {Farnell}},\ and\ \bibinfo {author} {\bibfnamefont {H.}~\bibnamefont {Rosner}},\ }\bibfield  {title} {\bibinfo {title} {Frustrated spin-$\frac{1}{2}$ ${J}_{1}\text{\ensuremath{-}}{J}_{2}$ {Heisenberg} ferromagnet on the square lattice studied via exact diagonalization and coupled-cluster method},\ }\href {https://doi.org/10.1103/PhysRevB.81.174429} {\bibfield  {journal} {\bibinfo  {journal} {Phys. Rev. B}\ }\textbf {\bibinfo {volume} {81}},\ \bibinfo {pages} {174429} (\bibinfo {year} {2010})}\BibitemShut {NoStop}%
\bibitem [{\citenamefont {Starykh}(2015)}]{Starykh2015}%
  \BibitemOpen
  \bibfield  {author} {\bibinfo {author} {\bibfnamefont {O.~A.}\ \bibnamefont {Starykh}},\ }\bibfield  {title} {\bibinfo {title} {{Unusual ordered phases of highly frustrated magnets: a review}},\ }\href {https://doi.org/10.1088/0034-4885/78/5/052502} {\bibfield  {journal} {\bibinfo  {journal} {Rep. Prog. Phys.}\ }\textbf {\bibinfo {volume} {78}},\ \bibinfo {pages} {052502} (\bibinfo {year} {2015})}\BibitemShut {NoStop}%
\bibitem [{\citenamefont {Jiang}\ \emph {et~al.}(2023{\natexlab{a}})\citenamefont {Jiang}, \citenamefont {Romh\'anyi}, \citenamefont {White}, \citenamefont {Zhitomirsky},\ and\ \citenamefont {Chernyshev}}]{Jiang2023QSN}%
  \BibitemOpen
  \bibfield  {author} {\bibinfo {author} {\bibfnamefont {S.}~\bibnamefont {Jiang}}, \bibinfo {author} {\bibfnamefont {J.}~\bibnamefont {Romh\'anyi}}, \bibinfo {author} {\bibfnamefont {S.~R.}\ \bibnamefont {White}}, \bibinfo {author} {\bibfnamefont {M.~E.}\ \bibnamefont {Zhitomirsky}},\ and\ \bibinfo {author} {\bibfnamefont {A.~L.}\ \bibnamefont {Chernyshev}},\ }\bibfield  {title} {\bibinfo {title} {{Where is the Quantum Spin Nematic?}},\ }\href {https://doi.org/10.1103/PhysRevLett.130.116701} {\bibfield  {journal} {\bibinfo  {journal} {Phys. Rev. Lett.}\ }\textbf {\bibinfo {volume} {130}},\ \bibinfo {pages} {116701} (\bibinfo {year} {2023}{\natexlab{a}})}\BibitemShut {NoStop}%
\bibitem [{\citenamefont {Jiang}\ \emph {et~al.}(2023{\natexlab{b}})\citenamefont {Jiang}, \citenamefont {White},\ and\ \citenamefont {Chernyshev}}]{jiang:2023b}%
  \BibitemOpen
  \bibfield  {author} {\bibinfo {author} {\bibfnamefont {S.}~\bibnamefont {Jiang}}, \bibinfo {author} {\bibfnamefont {S.~R.}\ \bibnamefont {White}},\ and\ \bibinfo {author} {\bibfnamefont {A.~L.}\ \bibnamefont {Chernyshev}},\ }\bibfield  {title} {\bibinfo {title} {Quantum phases in the honeycomb-lattice ${J}_{1}$--${J}_{3}$ ferro-antiferromagnetic model},\ }\href {https://doi.org/10.1103/PhysRevB.108.L180406} {\bibfield  {journal} {\bibinfo  {journal} {Phys. Rev. B}\ }\textbf {\bibinfo {volume} {108}},\ \bibinfo {pages} {L180406} (\bibinfo {year} {2023}{\natexlab{b}})}\BibitemShut {NoStop}%
\bibitem [{\citenamefont {Läuchli}\ \emph {et~al.}(2009)\citenamefont {Läuchli}, \citenamefont {Sudan},\ and\ \citenamefont {Lüscher}}]{lauchli:2009}%
  \BibitemOpen
  \bibfield  {author} {\bibinfo {author} {\bibfnamefont {A.~M.}\ \bibnamefont {Läuchli}}, \bibinfo {author} {\bibfnamefont {J.}~\bibnamefont {Sudan}},\ and\ \bibinfo {author} {\bibfnamefont {A.}~\bibnamefont {Lüscher}},\ }\bibfield  {title} {\bibinfo {title} {{The frustrated ferromagnetic $S=1/2$ {Heisenberg} chain in a magnetic field – How multipolar spin correlations emerge from magnetically ordered states}},\ }\href {https://doi.org/10.1088/1742-6596/145/1/012057} {\bibfield  {journal} {\bibinfo  {journal} {J. Phys.: Conf. Ser.}\ }\textbf {\bibinfo {volume} {145}},\ \bibinfo {pages} {012057} (\bibinfo {year} {2009})}\BibitemShut {NoStop}%
\bibitem [{\citenamefont {Momoi}\ \emph {et~al.}(2006)\citenamefont {Momoi}, \citenamefont {Sindzingre},\ and\ \citenamefont {Shannon}}]{momoi:2006}%
  \BibitemOpen
  \bibfield  {author} {\bibinfo {author} {\bibfnamefont {T.}~\bibnamefont {Momoi}}, \bibinfo {author} {\bibfnamefont {P.}~\bibnamefont {Sindzingre}},\ and\ \bibinfo {author} {\bibfnamefont {N.}~\bibnamefont {Shannon}},\ }\bibfield  {title} {\bibinfo {title} {{Octupolar Order in the Multiple Spin Exchange Model on a Triangular Lattice}},\ }\href {https://doi.org/10.1103/PhysRevLett.97.257204} {\bibfield  {journal} {\bibinfo  {journal} {Phys. Rev. Lett.}\ }\textbf {\bibinfo {volume} {97}},\ \bibinfo {pages} {257204} (\bibinfo {year} {2006})}\BibitemShut {NoStop}%
\bibitem [{\citenamefont {Derzhko}\ \emph {et~al.}(2015)\citenamefont {Derzhko}, \citenamefont {Richter},\ and\ \citenamefont {Maksymenko}}]{Derzhko:2015}%
  \BibitemOpen
  \bibfield  {author} {\bibinfo {author} {\bibfnamefont {O.}~\bibnamefont {Derzhko}}, \bibinfo {author} {\bibfnamefont {J.}~\bibnamefont {Richter}},\ and\ \bibinfo {author} {\bibfnamefont {M.}~\bibnamefont {Maksymenko}},\ }\bibfield  {title} {\bibinfo {title} {{Strongly correlated flat-band systems: The route from {Heisenberg} spins to Hubbard electrons}},\ }\href {https://doi.org/10.1142/S0217979215300078} {\bibfield  {journal} {\bibinfo  {journal} {Int. J. Mod. Phys. B}\ }\textbf {\bibinfo {volume} {29}},\ \bibinfo {pages} {1530007} (\bibinfo {year} {2015})}\BibitemShut {NoStop}%
\bibitem [{\citenamefont {Zhitomirsky}\ and\ \citenamefont {Tsunetsugu}(2004)}]{zhitomirsky:2004}%
  \BibitemOpen
  \bibfield  {author} {\bibinfo {author} {\bibfnamefont {M.~E.}\ \bibnamefont {Zhitomirsky}}\ and\ \bibinfo {author} {\bibfnamefont {H.}~\bibnamefont {Tsunetsugu}},\ }\bibfield  {title} {\bibinfo {title} {Exact low-temperature behavior of a kagom\'e antiferromagnet at high fields},\ }\href {https://doi.org/10.1103/PhysRevB.70.100403} {\bibfield  {journal} {\bibinfo  {journal} {Phys. Rev. B}\ }\textbf {\bibinfo {volume} {70}},\ \bibinfo {pages} {100403} (\bibinfo {year} {2004})}\BibitemShut {NoStop}%
\bibitem [{\citenamefont {Betts}(1995)}]{betts:1995}%
  \BibitemOpen
  \bibfield  {author} {\bibinfo {author} {\bibfnamefont {D.}~\bibnamefont {Betts}},\ }\bibfield  {title} {\bibinfo {title} {{A new two-dimensional lattice of coordination number five}},\ }\href {https://dalspace.library.dal.ca/items/354577a5-474d-4ddb-bfc5-862674ffbc0b} {\bibfield  {journal} {\bibinfo  {journal} {Proc. N. S. Inst. Sci.}\ }\textbf {\bibinfo {volume} {40}},\ \bibinfo {pages} {95} (\bibinfo {year} {1995})}\BibitemShut {NoStop}%
\bibitem [{\citenamefont {Sonnenschein}\ \emph {et~al.}(2024)\citenamefont {Sonnenschein}, \citenamefont {Maity}, \citenamefont {Liu}, \citenamefont {Thomale}, \citenamefont {Ferrari},\ and\ \citenamefont {Iqbal}}]{Sonnenschein2024}%
  \BibitemOpen
  \bibfield  {author} {\bibinfo {author} {\bibfnamefont {J.}~\bibnamefont {Sonnenschein}}, \bibinfo {author} {\bibfnamefont {A.}~\bibnamefont {Maity}}, \bibinfo {author} {\bibfnamefont {C.}~\bibnamefont {Liu}}, \bibinfo {author} {\bibfnamefont {R.}~\bibnamefont {Thomale}}, \bibinfo {author} {\bibfnamefont {F.}~\bibnamefont {Ferrari}},\ and\ \bibinfo {author} {\bibfnamefont {Y.}~\bibnamefont {Iqbal}},\ }\bibfield  {title} {\bibinfo {title} {{Candidate quantum spin liquids on the maple-leaf lattice}},\ }\href {https://doi.org/10.1103/PhysRevB.110.014414} {\bibfield  {journal} {\bibinfo  {journal} {Phys. Rev. B}\ }\textbf {\bibinfo {volume} {110}},\ \bibinfo {pages} {014414} (\bibinfo {year} {2024})}\BibitemShut {NoStop}%
\bibitem [{\citenamefont {Ghosh}\ \emph {et~al.}(2022)\citenamefont {Ghosh}, \citenamefont {M\"uller},\ and\ \citenamefont {Thomale}}]{Ghosh2022}%
  \BibitemOpen
  \bibfield  {author} {\bibinfo {author} {\bibfnamefont {P.}~\bibnamefont {Ghosh}}, \bibinfo {author} {\bibfnamefont {T.}~\bibnamefont {M\"uller}},\ and\ \bibinfo {author} {\bibfnamefont {R.}~\bibnamefont {Thomale}},\ }\bibfield  {title} {\bibinfo {title} {{Another exact ground state of a two-dimensional quantum antiferromagnet}},\ }\href {https://doi.org/10.1103/PhysRevB.105.L180412} {\bibfield  {journal} {\bibinfo  {journal} {Phys. Rev. B}\ }\textbf {\bibinfo {volume} {105}},\ \bibinfo {pages} {L180412} (\bibinfo {year} {2022})}\BibitemShut {NoStop}%
\bibitem [{\citenamefont {Farnell}\ \emph {et~al.}(2011)\citenamefont {Farnell}, \citenamefont {Darradi}, \citenamefont {Schmidt},\ and\ \citenamefont {Richter}}]{farnell:2011}%
  \BibitemOpen
  \bibfield  {author} {\bibinfo {author} {\bibfnamefont {D.~J.~J.}\ \bibnamefont {Farnell}}, \bibinfo {author} {\bibfnamefont {R.}~\bibnamefont {Darradi}}, \bibinfo {author} {\bibfnamefont {R.}~\bibnamefont {Schmidt}},\ and\ \bibinfo {author} {\bibfnamefont {J.}~\bibnamefont {Richter}},\ }\bibfield  {title} {\bibinfo {title} {{Spin-half {Heisenberg} antiferromagnet on two {A}rchimedian lattices: From the bounce lattice to the maple-leaf lattice and beyond}},\ }\href {https://doi.org/10.1103/PhysRevB.84.104406} {\bibfield  {journal} {\bibinfo  {journal} {Phys. Rev. B}\ }\textbf {\bibinfo {volume} {84}},\ \bibinfo {pages} {104406} (\bibinfo {year} {2011})}\BibitemShut {NoStop}%
\bibitem [{\citenamefont {Farnell}\ \emph {et~al.}(2014)\citenamefont {Farnell}, \citenamefont {G\"otze}, \citenamefont {Richter}, \citenamefont {Bishop},\ and\ \citenamefont {Li}}]{farnell:2014}%
  \BibitemOpen
  \bibfield  {author} {\bibinfo {author} {\bibfnamefont {D.~J.~J.}\ \bibnamefont {Farnell}}, \bibinfo {author} {\bibfnamefont {O.}~\bibnamefont {G\"otze}}, \bibinfo {author} {\bibfnamefont {J.}~\bibnamefont {Richter}}, \bibinfo {author} {\bibfnamefont {R.~F.}\ \bibnamefont {Bishop}},\ and\ \bibinfo {author} {\bibfnamefont {P.~H.~Y.}\ \bibnamefont {Li}},\ }\bibfield  {title} {\bibinfo {title} {{Quantum $s=\frac{1}{2}$ antiferromagnets on Archimedean lattices: The route from semiclassical magnetic order to nonmagnetic quantum states}},\ }\href {https://doi.org/10.1103/PhysRevB.89.184407} {\bibfield  {journal} {\bibinfo  {journal} {Phys. Rev. B}\ }\textbf {\bibinfo {volume} {89}},\ \bibinfo {pages} {184407} (\bibinfo {year} {2014})}\BibitemShut {NoStop}%
\bibitem [{\citenamefont {Beck}\ \emph {et~al.}(2024)\citenamefont {Beck}, \citenamefont {Bodky}, \citenamefont {Motruk}, \citenamefont {M{\"u}ller}, \citenamefont {Thomale},\ and\ \citenamefont {Ghosh}}]{beck:2024}%
  \BibitemOpen
  \bibfield  {author} {\bibinfo {author} {\bibfnamefont {J.}~\bibnamefont {Beck}}, \bibinfo {author} {\bibfnamefont {J.}~\bibnamefont {Bodky}}, \bibinfo {author} {\bibfnamefont {J.}~\bibnamefont {Motruk}}, \bibinfo {author} {\bibfnamefont {T.}~\bibnamefont {M{\"u}ller}}, \bibinfo {author} {\bibfnamefont {R.}~\bibnamefont {Thomale}},\ and\ \bibinfo {author} {\bibfnamefont {P.}~\bibnamefont {Ghosh}},\ }\bibfield  {title} {\bibinfo {title} {{Phase diagram of the \ensuremath{J-J_d} {Heisenberg} model on the maple leaf lattice: {Neural} networks and density matrix renormalization group}},\ }\href {https://doi.org/10.1103/PhysRevB.109.184422} {\bibfield  {journal} {\bibinfo  {journal} {Phys. Rev. B}\ }\textbf {\bibinfo {volume} {109}},\ \bibinfo {pages} {184422} (\bibinfo {year} {2024})}\BibitemShut {NoStop}%
\bibitem [{\citenamefont {Gresista}\ \emph {et~al.}(2023)\citenamefont {Gresista}, \citenamefont {Hickey}, \citenamefont {Trebst},\ and\ \citenamefont {Iqbal}}]{gresista:2023}%
  \BibitemOpen
  \bibfield  {author} {\bibinfo {author} {\bibfnamefont {L.}~\bibnamefont {Gresista}}, \bibinfo {author} {\bibfnamefont {C.}~\bibnamefont {Hickey}}, \bibinfo {author} {\bibfnamefont {S.}~\bibnamefont {Trebst}},\ and\ \bibinfo {author} {\bibfnamefont {Y.}~\bibnamefont {Iqbal}},\ }\bibfield  {title} {\bibinfo {title} {{Candidate quantum disordered intermediate phase in the {Heisenberg} antiferromagnet on the maple-leaf lattice}},\ }\href {https://doi.org/10.1103/PhysRevB.108.L241116} {\bibfield  {journal} {\bibinfo  {journal} {Phys. Rev. B}\ }\textbf {\bibinfo {volume} {108}},\ \bibinfo {pages} {L241116} (\bibinfo {year} {2023})}\BibitemShut {NoStop}%
\bibitem [{\citenamefont {Gresista}\ \emph {et~al.}(2026)\citenamefont {Gresista}, \citenamefont {Kiese}, \citenamefont {Trebst},\ and\ \citenamefont {Iqbal}}]{Gresista:2026}%
  \BibitemOpen
  \bibfield  {author} {\bibinfo {author} {\bibfnamefont {L.}~\bibnamefont {Gresista}}, \bibinfo {author} {\bibfnamefont {D.}~\bibnamefont {Kiese}}, \bibinfo {author} {\bibfnamefont {S.}~\bibnamefont {Trebst}},\ and\ \bibinfo {author} {\bibfnamefont {Y.}~\bibnamefont {Iqbal}},\ }\bibfield  {title} {\bibinfo {title} {{Unconventional orders in the maple-leaf ferro-antiferromagnetic {Heisenberg} model}},\ }\href {https://doi.org/10.1515/zna-2025-0376} {\bibfield  {journal} {\bibinfo  {journal} {Z. Naturforsch. A}\ } (\bibinfo {year} {2026})}\BibitemShut {NoStop}%
\bibitem [{\citenamefont {Gemb\'e}\ \emph {et~al.}(2024)\citenamefont {Gemb\'e}, \citenamefont {Gresista}, \citenamefont {Schmidt}, \citenamefont {Hickey}, \citenamefont {Iqbal},\ and\ \citenamefont {Trebst}}]{gembe:2024}%
  \BibitemOpen
  \bibfield  {author} {\bibinfo {author} {\bibfnamefont {M.}~\bibnamefont {Gemb\'e}}, \bibinfo {author} {\bibfnamefont {L.}~\bibnamefont {Gresista}}, \bibinfo {author} {\bibfnamefont {H.-J.}\ \bibnamefont {Schmidt}}, \bibinfo {author} {\bibfnamefont {C.}~\bibnamefont {Hickey}}, \bibinfo {author} {\bibfnamefont {Y.}~\bibnamefont {Iqbal}},\ and\ \bibinfo {author} {\bibfnamefont {S.}~\bibnamefont {Trebst}},\ }\bibfield  {title} {\bibinfo {title} {{Noncoplanar orders and quantum disordered states in maple-leaf antiferromagnets}},\ }\href {https://doi.org/10.1103/PhysRevB.110.085151} {\bibfield  {journal} {\bibinfo  {journal} {Phys. Rev. B}\ }\textbf {\bibinfo {volume} {110}},\ \bibinfo {pages} {085151} (\bibinfo {year} {2024})}\BibitemShut {NoStop}%
\bibitem [{\citenamefont {Schmoll}\ \emph {et~al.}(2025)\citenamefont {Schmoll}, \citenamefont {Naumann}, \citenamefont {Weerda}, \citenamefont {Eisert},\ and\ \citenamefont {Iqbal}}]{schmoll:2025}%
  \BibitemOpen
  \bibfield  {author} {\bibinfo {author} {\bibfnamefont {P.}~\bibnamefont {Schmoll}}, \bibinfo {author} {\bibfnamefont {J.}~\bibnamefont {Naumann}}, \bibinfo {author} {\bibfnamefont {E.~L.}\ \bibnamefont {Weerda}}, \bibinfo {author} {\bibfnamefont {J.}~\bibnamefont {Eisert}},\ and\ \bibinfo {author} {\bibfnamefont {Y.}~\bibnamefont {Iqbal}},\ }\href {https://arxiv.org/abs/2407.07145} {\bibinfo {title} {{Bathing in a sea of candidate quantum spin liquids: From the gapless ruby to the gapped maple-leaf lattice}}} (\bibinfo {year} {2025}),\ \Eprint {https://arxiv.org/abs/2407.07145} {arXiv:2407.07145 [cond-mat.str-el]} \BibitemShut {NoStop}%
\bibitem [{\citenamefont {Ebert}\ \emph {et~al.}(2026)\citenamefont {Ebert}, \citenamefont {Iqbal},\ and\ \citenamefont {Wietek}}]{Ebert2026}%
  \BibitemOpen
  \bibfield  {author} {\bibinfo {author} {\bibfnamefont {P.~L.}\ \bibnamefont {Ebert}}, \bibinfo {author} {\bibfnamefont {Y.}~\bibnamefont {Iqbal}},\ and\ \bibinfo {author} {\bibfnamefont {A.}~\bibnamefont {Wietek}},\ }\bibfield  {title} {\bibinfo {title} {{Competing Paramagnetic Phases in the Maple-Leaf {Heisenberg} Antiferromagnet}},\ }\href@noop {} {\bibfield  {journal} {\bibinfo  {journal} {arXiv}\ } (\bibinfo {year} {2026})},\ \Eprint {https://arxiv.org/abs/2601.05308} {2601.05308} \BibitemShut {NoStop}%
\bibitem [{\citenamefont {Ghosh}(2025)}]{ghosh:2025b}%
  \BibitemOpen
  \bibfield  {author} {\bibinfo {author} {\bibfnamefont {P.}~\bibnamefont {Ghosh}},\ }\bibfield  {title} {\bibinfo {title} {Chiral crossroads in {Ho}$_3${Sc}{O}$_6$: Competing interactions on the maple-leaf lattice},\ }\href {https://doi.org/10.1103/3zlw-hrbf} {\bibfield  {journal} {\bibinfo  {journal} {Phys. Rev. B}\ }\textbf {\bibinfo {volume} {111}},\ \bibinfo {pages} {224431} (\bibinfo {year} {2025})}\BibitemShut {NoStop}%
\bibitem [{\citenamefont {Aguilar-Maldonado}\ \emph {et~al.}(2025)\citenamefont {Aguilar-Maldonado}, \citenamefont {Feyerherm}, \citenamefont {Proke\ifmmode~\check{s}\else \v{s}\fi{}}, \citenamefont {Keller},\ and\ \citenamefont {Lake}}]{Aguilar-Maldonado:2025}%
  \BibitemOpen
  \bibfield  {author} {\bibinfo {author} {\bibfnamefont {C.}~\bibnamefont {Aguilar-Maldonado}}, \bibinfo {author} {\bibfnamefont {R.}~\bibnamefont {Feyerherm}}, \bibinfo {author} {\bibfnamefont {K.}~\bibnamefont {Proke\ifmmode~\check{s}\else \v{s}\fi{}}}, \bibinfo {author} {\bibfnamefont {L.}~\bibnamefont {Keller}},\ and\ \bibinfo {author} {\bibfnamefont {B.}~\bibnamefont {Lake}},\ }\bibfield  {title} {\bibinfo {title} {Structure and magnetic properties of the maple leaf antiferromagnet {Ho}$_{3}${ScO}$_{6}$},\ }\href {https://doi.org/10.1103/PhysRevB.111.094439} {\bibfield  {journal} {\bibinfo  {journal} {Phys. Rev. B}\ }\textbf {\bibinfo {volume} {111}},\ \bibinfo {pages} {094439} (\bibinfo {year} {2025})}\BibitemShut {NoStop}%
\bibitem [{\citenamefont {Haraguchi}\ \emph {et~al.}(2018)\citenamefont {Haraguchi}, \citenamefont {Matsuo}, \citenamefont {Kindo},\ and\ \citenamefont {Hiroi}}]{Haraguchi2018}%
  \BibitemOpen
  \bibfield  {author} {\bibinfo {author} {\bibfnamefont {Y.}~\bibnamefont {Haraguchi}}, \bibinfo {author} {\bibfnamefont {A.}~\bibnamefont {Matsuo}}, \bibinfo {author} {\bibfnamefont {K.}~\bibnamefont {Kindo}},\ and\ \bibinfo {author} {\bibfnamefont {Z.}~\bibnamefont {Hiroi}},\ }\bibfield  {title} {\bibinfo {title} {{Frustrated magnetism of the maple-leaf-lattice antiferromagnet MgMn$_3$O$_7$$\cdot$3H$_2$O}},\ }\href {https://doi.org/10.1103/PhysRevB.98.064412} {\bibfield  {journal} {\bibinfo  {journal} {Phys. Rev. B}\ }\textbf {\bibinfo {volume} {98}},\ \bibinfo {pages} {064412} (\bibinfo {year} {2018})}\BibitemShut {NoStop}%
\bibitem [{\citenamefont {Saha}\ \emph {et~al.}(2023)\citenamefont {Saha}, \citenamefont {Bera}, \citenamefont {Yusuf},\ and\ \citenamefont {Hoser}}]{saha:2023}%
  \BibitemOpen
  \bibfield  {author} {\bibinfo {author} {\bibfnamefont {B.}~\bibnamefont {Saha}}, \bibinfo {author} {\bibfnamefont {A.~K.}\ \bibnamefont {Bera}}, \bibinfo {author} {\bibfnamefont {S.~M.}\ \bibnamefont {Yusuf}},\ and\ \bibinfo {author} {\bibfnamefont {A.}~\bibnamefont {Hoser}},\ }\bibfield  {title} {\bibinfo {title} {{Two-dimensional short-range spin-spin correlations in the layered spin-$\frac{3}{2}$ maple leaf lattice antiferromagnet {Na}$_{2}${Mn}$_{3}${O}$_{7}$ with crystal stacking disorder}},\ }\href {https://doi.org/10.1103/PhysRevB.107.064419} {\bibfield  {journal} {\bibinfo  {journal} {Phys. Rev. B}\ }\textbf {\bibinfo {volume} {107}},\ \bibinfo {pages} {064419} (\bibinfo {year} {2023})}\BibitemShut {NoStop}%
\bibitem [{\citenamefont {Nakano}\ and\ \citenamefont {Sakai}(2026)}]{nakano:2026}%
  \BibitemOpen
  \bibfield  {author} {\bibinfo {author} {\bibfnamefont {H.}~\bibnamefont {Nakano}}\ and\ \bibinfo {author} {\bibfnamefont {T.}~\bibnamefont {Sakai}},\ }\href {https://arxiv.org/abs/2604.24055} {\bibinfo {title} {Spin excitation of the {Heisenberg} antiferromagnet with frustration: from the bounce-lattice antiferromagnet through the maple-leaf-lattice antiferromagnet to the exact-dimer system}} (\bibinfo {year} {2026}),\ \Eprint {https://arxiv.org/abs/2604.24055} {arXiv:2604.24055 [cond-mat.mtrl-sci]} \BibitemShut {NoStop}%
\bibitem [{\citenamefont {Ghosh}\ \emph {et~al.}(2023)\citenamefont {Ghosh}, \citenamefont {Seufert}, \citenamefont {M\"uller}, \citenamefont {Mila},\ and\ \citenamefont {Thomale}}]{Ghosh2023Field}%
  \BibitemOpen
  \bibfield  {author} {\bibinfo {author} {\bibfnamefont {P.}~\bibnamefont {Ghosh}}, \bibinfo {author} {\bibfnamefont {J.}~\bibnamefont {Seufert}}, \bibinfo {author} {\bibfnamefont {T.}~\bibnamefont {M\"uller}}, \bibinfo {author} {\bibfnamefont {F.}~\bibnamefont {Mila}},\ and\ \bibinfo {author} {\bibfnamefont {R.}~\bibnamefont {Thomale}},\ }\bibfield  {title} {\bibinfo {title} {{Maple leaf antiferromagnet in a magnetic field}},\ }\href {https://doi.org/10.1103/PhysRevB.108.L060406} {\bibfield  {journal} {\bibinfo  {journal} {Phys. Rev. B}\ }\textbf {\bibinfo {volume} {108}},\ \bibinfo {pages} {L060406} (\bibinfo {year} {2023})}\BibitemShut {NoStop}%
\bibitem [{\citenamefont {Ghosh}\ \emph {et~al.}(2024)\citenamefont {Ghosh}, \citenamefont {Müller}, \citenamefont {Iqbal}, \citenamefont {Thomale},\ and\ \citenamefont {Jeschke}}]{ghosh:2024c}%
  \BibitemOpen
  \bibfield  {author} {\bibinfo {author} {\bibfnamefont {P.}~\bibnamefont {Ghosh}}, \bibinfo {author} {\bibfnamefont {T.}~\bibnamefont {Müller}}, \bibinfo {author} {\bibfnamefont {Y.}~\bibnamefont {Iqbal}}, \bibinfo {author} {\bibfnamefont {R.}~\bibnamefont {Thomale}},\ and\ \bibinfo {author} {\bibfnamefont {H.~O.}\ \bibnamefont {Jeschke}},\ }\bibfield  {title} {\bibinfo {title} {Effective spin-1 breathing kagome {Hamiltonian} induced by the exchange hierarchy in the maple leaf mineral bluebellite},\ }\href {https://doi.org/10.1103/PhysRevB.110.094406} {\bibfield  {journal} {\bibinfo  {journal} {Phys. Rev. B}\ }\textbf {\bibinfo {volume} {110}},\ \bibinfo {pages} {094406} (\bibinfo {year} {2024})},\ \bibinfo {note} {publisher: American Physical Society}\BibitemShut {NoStop}%
\bibitem [{\citenamefont {Schmoll}\ \emph {et~al.}()\citenamefont {Schmoll}, \citenamefont {Jeschke},\ and\ \citenamefont {Iqbal}}]{schmoll:2024a}%
  \BibitemOpen
  \bibfield  {author} {\bibinfo {author} {\bibfnamefont {P.}~\bibnamefont {Schmoll}}, \bibinfo {author} {\bibfnamefont {H.}~\bibnamefont {Jeschke}},\ and\ \bibinfo {author} {\bibfnamefont {Y.}~\bibnamefont {Iqbal}},\ }\bibfield  {title} {\bibinfo {title} {Tensor network analysis of the maple-leaf antiferromagnet spangolite},\ }\href {https://doi.org/10.1038/s43246-025-00904-1} {\bibfield  {journal} {\bibinfo  {journal} {Commun Mater}\ }\textbf {\bibinfo {volume} {6}}}\BibitemShut {NoStop}%
\bibitem [{\citenamefont {Venkatesh}\ \emph {et~al.}(2020)\citenamefont {Venkatesh}, \citenamefont {Bandyopadhyay}, \citenamefont {Midya}, \citenamefont {Mahalingam}, \citenamefont {Ganesan},\ and\ \citenamefont {Mandal}}]{venkatesh:2020}%
  \BibitemOpen
  \bibfield  {author} {\bibinfo {author} {\bibfnamefont {C.}~\bibnamefont {Venkatesh}}, \bibinfo {author} {\bibfnamefont {B.}~\bibnamefont {Bandyopadhyay}}, \bibinfo {author} {\bibfnamefont {A.}~\bibnamefont {Midya}}, \bibinfo {author} {\bibfnamefont {K.}~\bibnamefont {Mahalingam}}, \bibinfo {author} {\bibfnamefont {V.}~\bibnamefont {Ganesan}},\ and\ \bibinfo {author} {\bibfnamefont {P.}~\bibnamefont {Mandal}},\ }\bibfield  {title} {\bibinfo {title} {{Magnetic properties of the one-dimensional $S=\frac{3}{2}$ {Heisenberg} antiferromagnetic spin-chain compound {Na}$_{2}${Mn}$_{3}${O}$_{7}$}},\ }\href {https://doi.org/10.1103/PhysRevB.101.184429} {\bibfield  {journal} {\bibinfo  {journal} {Phys. Rev. B}\ }\textbf {\bibinfo {volume} {101}},\ \bibinfo {pages} {184429} (\bibinfo {year} {2020})}\BibitemShut {NoStop}%
\bibitem [{\citenamefont {Haraguchi}\ \emph {et~al.}(2021)\citenamefont {Haraguchi}, \citenamefont {Matsuo}, \citenamefont {Kindo},\ and\ \citenamefont {Hiroi}}]{haraguchi:2021}%
  \BibitemOpen
  \bibfield  {author} {\bibinfo {author} {\bibfnamefont {Y.}~\bibnamefont {Haraguchi}}, \bibinfo {author} {\bibfnamefont {A.}~\bibnamefont {Matsuo}}, \bibinfo {author} {\bibfnamefont {K.}~\bibnamefont {Kindo}},\ and\ \bibinfo {author} {\bibfnamefont {Z.}~\bibnamefont {Hiroi}},\ }\bibfield  {title} {\bibinfo {title} {{Quantum antiferromagnet bluebellite comprising a maple-leaf lattice made of spin-1/2 {C}u$^{2+}$ ions}},\ }\href {https://doi.org/10.1103/PhysRevB.104.174439} {\bibfield  {journal} {\bibinfo  {journal} {Phys. Rev. B}\ }\textbf {\bibinfo {volume} {104}},\ \bibinfo {pages} {174439} (\bibinfo {year} {2021})}\BibitemShut {NoStop}%
\bibitem [{\citenamefont {Sch{\"a}fer}\ \emph {et~al.}(2026)\citenamefont {Sch{\"a}fer}, \citenamefont {Ebert}, \citenamefont {Hassan}, \citenamefont {Reuther}, \citenamefont {Luitz},\ and\ \citenamefont {Wietek}}]{Schaefer:2026}%
  \BibitemOpen
  \bibfield  {author} {\bibinfo {author} {\bibfnamefont {R.}~\bibnamefont {Sch{\"a}fer}}, \bibinfo {author} {\bibfnamefont {P.~L.}\ \bibnamefont {Ebert}}, \bibinfo {author} {\bibfnamefont {N.}~\bibnamefont {Hassan}}, \bibinfo {author} {\bibfnamefont {J.}~\bibnamefont {Reuther}}, \bibinfo {author} {\bibfnamefont {D.~J.}\ \bibnamefont {Luitz}},\ and\ \bibinfo {author} {\bibfnamefont {A.}~\bibnamefont {Wietek}},\ }\bibfield  {title} {\bibinfo {title} {{Thermodynamics of the {Heisenberg} antiferromagnet on the maple-leaf lattice}},\ }\href {https://doi.org/10.1515/zna-2025-0382} {\bibfield  {journal} {\bibinfo  {journal} {Z. Naturforsch. A}\ } (\bibinfo {year} {2026})}\BibitemShut {NoStop}%
\bibitem [{\citenamefont {Borutta}\ \emph {et~al.}(2026)\citenamefont {Borutta}, \citenamefont {Müller}, \citenamefont {Thomale}, \citenamefont {Jeschke},\ and\ \citenamefont {Iqbal}}]{borutta:2026}%
  \BibitemOpen
  \bibfield  {author} {\bibinfo {author} {\bibfnamefont {H.}~\bibnamefont {Borutta}}, \bibinfo {author} {\bibfnamefont {T.}~\bibnamefont {Müller}}, \bibinfo {author} {\bibfnamefont {R.}~\bibnamefont {Thomale}}, \bibinfo {author} {\bibfnamefont {H.~O.}\ \bibnamefont {Jeschke}},\ and\ \bibinfo {author} {\bibfnamefont {Y.}~\bibnamefont {Iqbal}},\ }\href {https://arxiv.org/abs/2602.22005} {\bibinfo {title} {Crystallography-driven molecularization of a two-dimensional spin-$3/2$ magnet}} (\bibinfo {year} {2026}),\ \Eprint {https://arxiv.org/abs/2602.22005} {arXiv:2602.22005 [cond-mat.str-el]} \BibitemShut {NoStop}%
\bibitem [{\citenamefont {Ghosh}(2024)}]{ghosh:2024a}%
  \BibitemOpen
  \bibfield  {author} {\bibinfo {author} {\bibfnamefont {P.}~\bibnamefont {Ghosh}},\ }\href {https://arxiv.org/abs/2401.09422} {\bibinfo {title} {Where is the spin liquid in maple-leaf quantum magnet?}} (\bibinfo {year} {2024}),\ \Eprint {https://arxiv.org/abs/2401.09422} {arXiv:2401.09422 [cond-mat.str-el]} \BibitemShut {NoStop}%
\bibitem [{\citenamefont {Mills}\ \emph {et~al.}(2014)\citenamefont {Mills}, \citenamefont {Kampf}, \citenamefont {Christy}, \citenamefont {Housley}, \citenamefont {Rossman}, \citenamefont {Reynolds},\ and\ \citenamefont {Marty}}]{mills:2014}%
  \BibitemOpen
  \bibfield  {author} {\bibinfo {author} {\bibfnamefont {S.~J.}\ \bibnamefont {Mills}}, \bibinfo {author} {\bibfnamefont {A.~R.}\ \bibnamefont {Kampf}}, \bibinfo {author} {\bibfnamefont {A.~G.}\ \bibnamefont {Christy}}, \bibinfo {author} {\bibfnamefont {R.~M.}\ \bibnamefont {Housley}}, \bibinfo {author} {\bibfnamefont {G.~R.}\ \bibnamefont {Rossman}}, \bibinfo {author} {\bibfnamefont {R.~E.}\ \bibnamefont {Reynolds}},\ and\ \bibinfo {author} {\bibfnamefont {J.}~\bibnamefont {Marty}},\ }\bibfield  {title} {\bibinfo {title} {Bluebellite and {M}ojaveite, two new minerals from the central {M}ojave {D}esert, {C}alifornia, {USA}},\ }\href {https://doi.org/10.1180/minmag.2014.078.5.15} {\bibfield  {journal} {\bibinfo  {journal} {Mineralog. Mag.}\ }\textbf {\bibinfo {volume} {78}},\ \bibinfo {pages} {1325–1340} (\bibinfo {year} {2014})}\BibitemShut {NoStop}%
\bibitem [{\citenamefont {Fennell}\ \emph {et~al.}(2011)\citenamefont {Fennell}, \citenamefont {Piatek}, \citenamefont {Stephenson}, \citenamefont {Nilsen},\ and\ \citenamefont {Rønnow}}]{fennell:2011}%
  \BibitemOpen
  \bibfield  {author} {\bibinfo {author} {\bibfnamefont {T.}~\bibnamefont {Fennell}}, \bibinfo {author} {\bibfnamefont {J.~O.}\ \bibnamefont {Piatek}}, \bibinfo {author} {\bibfnamefont {R.~A.}\ \bibnamefont {Stephenson}}, \bibinfo {author} {\bibfnamefont {G.~J.}\ \bibnamefont {Nilsen}},\ and\ \bibinfo {author} {\bibfnamefont {H.~M.}\ \bibnamefont {Rønnow}},\ }\bibfield  {title} {\bibinfo {title} {Spangolite: an s = 1/2 maple leaf lattice antiferromagnet?},\ }\href {https://doi.org/10.1088/0953-8984/23/16/164201} {\bibfield  {journal} {\bibinfo  {journal} {J. Phys.: Condens. Matter}\ }\textbf {\bibinfo {volume} {23}},\ \bibinfo {pages} {164201} (\bibinfo {year} {2011})}\BibitemShut {NoStop}%
\bibitem [{\citenamefont {Wietek}\ \emph {et~al.}(2026{\natexlab{a}})\citenamefont {Wietek}, \citenamefont {Staszewski}, \citenamefont {Ulaga}, \citenamefont {Ebert}, \citenamefont {Karlsson}, \citenamefont {Sarkar}, \citenamefont {Shackleton}, \citenamefont {Sinha},\ and\ \citenamefont {Soares}}]{wietek:2025:xdiag:paper}%
  \BibitemOpen
  \bibfield  {author} {\bibinfo {author} {\bibfnamefont {A.}~\bibnamefont {Wietek}}, \bibinfo {author} {\bibfnamefont {L.}~\bibnamefont {Staszewski}}, \bibinfo {author} {\bibfnamefont {M.}~\bibnamefont {Ulaga}}, \bibinfo {author} {\bibfnamefont {P.~L.}\ \bibnamefont {Ebert}}, \bibinfo {author} {\bibfnamefont {H.}~\bibnamefont {Karlsson}}, \bibinfo {author} {\bibfnamefont {S.}~\bibnamefont {Sarkar}}, \bibinfo {author} {\bibfnamefont {L.}~\bibnamefont {Shackleton}}, \bibinfo {author} {\bibfnamefont {A.}~\bibnamefont {Sinha}},\ and\ \bibinfo {author} {\bibfnamefont {R.~D.}\ \bibnamefont {Soares}},\ }\bibfield  {title} {\bibinfo {title} {{XDiag: Exact diagonalization for quantum many-body systems}},\ }\href {https://doi.org/10.21468/SciPostPhysCodeb.70} {\bibfield  {journal} {\bibinfo  {journal} {SciPost Phys. Codebases}\ ,\ \bibinfo {pages} {70}} (\bibinfo {year} {2026}{\natexlab{a}})}\BibitemShut {NoStop}%
\bibitem [{\citenamefont {Wietek}\ \emph {et~al.}(2026{\natexlab{b}})\citenamefont {Wietek}, \citenamefont {Staszewski}, \citenamefont {Ulaga}, \citenamefont {Ebert}, \citenamefont {Karlsson}, \citenamefont {Sarkar}, \citenamefont {Shackleton}, \citenamefont {Sinha},\ and\ \citenamefont {Soares}}]{wietek:2025:xdiag:code}%
  \BibitemOpen
  \bibfield  {author} {\bibinfo {author} {\bibfnamefont {A.}~\bibnamefont {Wietek}}, \bibinfo {author} {\bibfnamefont {L.}~\bibnamefont {Staszewski}}, \bibinfo {author} {\bibfnamefont {M.}~\bibnamefont {Ulaga}}, \bibinfo {author} {\bibfnamefont {P.~L.}\ \bibnamefont {Ebert}}, \bibinfo {author} {\bibfnamefont {H.}~\bibnamefont {Karlsson}}, \bibinfo {author} {\bibfnamefont {S.}~\bibnamefont {Sarkar}}, \bibinfo {author} {\bibfnamefont {L.}~\bibnamefont {Shackleton}}, \bibinfo {author} {\bibfnamefont {A.}~\bibnamefont {Sinha}},\ and\ \bibinfo {author} {\bibfnamefont {R.~D.}\ \bibnamefont {Soares}},\ }\bibfield  {title} {\bibinfo {title} {{Codebase release 0.4 for XDiag}},\ }\href {https://doi.org/10.21468/SciPostPhysCodeb.70-r0.4} {\bibfield  {journal} {\bibinfo  {journal} {SciPost Phys. Codebases}\ ,\ \bibinfo {pages} {70}} (\bibinfo {year} {2026}{\natexlab{b}})}\BibitemShut {NoStop}%
\bibitem [{Sup()}]{SupplementalMaterial}%
  \BibitemOpen
  \href@noop {} {\bibinfo {title} {{See Supplemental Material at [URL will be inserted by publisher] for two-magnon interaction data, ground state energies over $S$, the employed finite size clusters and a table of all surveyed points and ground state energies.}}}\BibitemShut {Stop}%
\bibitem [{Note1()}]{Note1}%
  \BibitemOpen
  \bibinfo {note} {Recall that the maple-leaf lattice can be obtained by a one-seventh site depletion of the triangular lattice.}\BibitemShut {Stop}%
\bibitem [{\citenamefont {Wietek}\ and\ \citenamefont {L\"auchli}(2020)}]{wietek:2020}%
  \BibitemOpen
  \bibfield  {author} {\bibinfo {author} {\bibfnamefont {A.}~\bibnamefont {Wietek}}\ and\ \bibinfo {author} {\bibfnamefont {A.~M.}\ \bibnamefont {L\"auchli}},\ }\bibfield  {title} {\bibinfo {title} {{Valence bond solid and possible deconfined quantum criticality in an extended kagome lattice {Heisenberg} antiferromagnet}},\ }\href {https://doi.org/10.1103/PhysRevB.102.020411} {\bibfield  {journal} {\bibinfo  {journal} {Phys. Rev. B}\ }\textbf {\bibinfo {volume} {102}},\ \bibinfo {pages} {020411} (\bibinfo {year} {2020})}\BibitemShut {NoStop}%
\bibitem [{\citenamefont {Penc}\ and\ \citenamefont {L{\"a}uchli}(2011)}]{penc:2011}%
  \BibitemOpen
  \bibfield  {author} {\bibinfo {author} {\bibfnamefont {K.}~\bibnamefont {Penc}}\ and\ \bibinfo {author} {\bibfnamefont {A.~M.}\ \bibnamefont {L{\"a}uchli}},\ }\bibinfo {title} {{Spin Nematic Phases in Quantum Spin Systems}},\ in\ \href {https://doi.org/10.1007/978-3-642-10589-0_13} {\emph {\bibinfo {booktitle} {Introduction to Frustrated Magnetism: Materials, Experiments, Theory}}},\ \bibinfo {editor} {edited by\ \bibinfo {editor} {\bibfnamefont {C.}~\bibnamefont {Lacroix}}, \bibinfo {editor} {\bibfnamefont {P.}~\bibnamefont {Mendels}},\ and\ \bibinfo {editor} {\bibfnamefont {F.}~\bibnamefont {Mila}}}\ (\bibinfo  {publisher} {Springer Berlin Heidelberg},\ \bibinfo {address} {Berlin, Heidelberg},\ \bibinfo {year} {2011})\ pp.\ \bibinfo {pages} {331--362}\BibitemShut {NoStop}%
\bibitem [{\citenamefont {Iqbal}\ \emph {et~al.}(2013)\citenamefont {Iqbal}, \citenamefont {Becca}, \citenamefont {Sorella},\ and\ \citenamefont {Poilblanc}}]{Iqbal-2013}%
  \BibitemOpen
  \bibfield  {author} {\bibinfo {author} {\bibfnamefont {Y.}~\bibnamefont {Iqbal}}, \bibinfo {author} {\bibfnamefont {F.}~\bibnamefont {Becca}}, \bibinfo {author} {\bibfnamefont {S.}~\bibnamefont {Sorella}},\ and\ \bibinfo {author} {\bibfnamefont {D.}~\bibnamefont {Poilblanc}},\ }\bibfield  {title} {\bibinfo {title} {{Gapless spin-liquid phase in the kagome spin-$\frac{1}{2}$ {Heisenberg} antiferromagnet}},\ }\href {https://doi.org/10.1103/PhysRevB.87.060405} {\bibfield  {journal} {\bibinfo  {journal} {Phys. Rev. B}\ }\textbf {\bibinfo {volume} {87}},\ \bibinfo {pages} {060405} (\bibinfo {year} {2013})}\BibitemShut {NoStop}%
\bibitem [{\citenamefont {Wietek}\ \emph {et~al.}(2024)\citenamefont {Wietek}, \citenamefont {Capponi},\ and\ \citenamefont {L\"auchli}}]{wietek:2024}%
  \BibitemOpen
  \bibfield  {author} {\bibinfo {author} {\bibfnamefont {A.}~\bibnamefont {Wietek}}, \bibinfo {author} {\bibfnamefont {S.}~\bibnamefont {Capponi}},\ and\ \bibinfo {author} {\bibfnamefont {A.~M.}\ \bibnamefont {L\"auchli}},\ }\bibfield  {title} {\bibinfo {title} {{Quantum Electrodynamics in $2+1$ Dimensions as the Organizing Principle of a Triangular Lattice Antiferromagnet}},\ }\href {https://doi.org/10.1103/PhysRevX.14.021010} {\bibfield  {journal} {\bibinfo  {journal} {Phys. Rev. X}\ }\textbf {\bibinfo {volume} {14}},\ \bibinfo {pages} {021010} (\bibinfo {year} {2024})}\BibitemShut {NoStop}%
\bibitem [{\citenamefont {Semeghini}\ \emph {et~al.}(2021)\citenamefont {Semeghini}, \citenamefont {Levine}, \citenamefont {Keesling}, \citenamefont {Ebadi}, \citenamefont {Wang}, \citenamefont {Bluvstein}, \citenamefont {Verresen}, \citenamefont {Pichler}, \citenamefont {Kalinowski}, \citenamefont {Samajdar}, \citenamefont {Omran}, \citenamefont {Sachdev}, \citenamefont {Vishwanath}, \citenamefont {Greiner}, \citenamefont {Vuleti{\'c}},\ and\ \citenamefont {Lukin}}]{semeghini:2021}%
  \BibitemOpen
  \bibfield  {author} {\bibinfo {author} {\bibfnamefont {G.}~\bibnamefont {Semeghini}}, \bibinfo {author} {\bibfnamefont {H.}~\bibnamefont {Levine}}, \bibinfo {author} {\bibfnamefont {A.}~\bibnamefont {Keesling}}, \bibinfo {author} {\bibfnamefont {S.}~\bibnamefont {Ebadi}}, \bibinfo {author} {\bibfnamefont {T.~T.}\ \bibnamefont {Wang}}, \bibinfo {author} {\bibfnamefont {D.}~\bibnamefont {Bluvstein}}, \bibinfo {author} {\bibfnamefont {R.}~\bibnamefont {Verresen}}, \bibinfo {author} {\bibfnamefont {H.}~\bibnamefont {Pichler}}, \bibinfo {author} {\bibfnamefont {M.}~\bibnamefont {Kalinowski}}, \bibinfo {author} {\bibfnamefont {R.}~\bibnamefont {Samajdar}}, \bibinfo {author} {\bibfnamefont {A.}~\bibnamefont {Omran}}, \bibinfo {author} {\bibfnamefont {S.}~\bibnamefont {Sachdev}}, \bibinfo {author} {\bibfnamefont {A.}~\bibnamefont {Vishwanath}}, \bibinfo {author} {\bibfnamefont {M.}~\bibnamefont {Greiner}}, \bibinfo {author} {\bibfnamefont {V.}~\bibnamefont {Vuleti{\'c}}},\ and\ \bibinfo {author} {\bibfnamefont
  {M.~D.}\ \bibnamefont {Lukin}},\ }\bibfield  {title} {\bibinfo {title} {{Probing topological spin liquids on a programmable quantum simulator}},\ }\href {https://doi.org/10.1126/science.abi8794} {\bibfield  {journal} {\bibinfo  {journal} {Science}\ }\textbf {\bibinfo {volume} {374}},\ \bibinfo {pages} {1242} (\bibinfo {year} {2021})}\BibitemShut {NoStop}%
\bibitem [{\citenamefont {Giudici}\ \emph {et~al.}(2022)\citenamefont {Giudici}, \citenamefont {Lukin},\ and\ \citenamefont {Pichler}}]{giudici:2022}%
  \BibitemOpen
  \bibfield  {author} {\bibinfo {author} {\bibfnamefont {G.}~\bibnamefont {Giudici}}, \bibinfo {author} {\bibfnamefont {M.~D.}\ \bibnamefont {Lukin}},\ and\ \bibinfo {author} {\bibfnamefont {H.}~\bibnamefont {Pichler}},\ }\bibfield  {title} {\bibinfo {title} {{Dynamical Preparation of Quantum Spin Liquids in Rydberg Atom Arrays}},\ }\href {https://doi.org/10.1103/PhysRevLett.129.090401} {\bibfield  {journal} {\bibinfo  {journal} {Phys. Rev. Lett.}\ }\textbf {\bibinfo {volume} {129}},\ \bibinfo {pages} {090401} (\bibinfo {year} {2022})}\BibitemShut {NoStop}%
\bibitem [{\citenamefont {Samajdar}\ \emph {et~al.}(2023)\citenamefont {Samajdar}, \citenamefont {Joshi}, \citenamefont {Teng},\ and\ \citenamefont {Sachdev}}]{samajdar:2023}%
  \BibitemOpen
  \bibfield  {author} {\bibinfo {author} {\bibfnamefont {R.}~\bibnamefont {Samajdar}}, \bibinfo {author} {\bibfnamefont {D.~G.}\ \bibnamefont {Joshi}}, \bibinfo {author} {\bibfnamefont {Y.}~\bibnamefont {Teng}},\ and\ \bibinfo {author} {\bibfnamefont {S.}~\bibnamefont {Sachdev}},\ }\bibfield  {title} {\bibinfo {title} {{Emergent ${\mathbb{Z}}_{2}$ Gauge Theories and Topological Excitations in Rydberg Atom Arrays}},\ }\href {https://doi.org/10.1103/PhysRevLett.130.043601} {\bibfield  {journal} {\bibinfo  {journal} {Phys. Rev. Lett.}\ }\textbf {\bibinfo {volume} {130}},\ \bibinfo {pages} {043601} (\bibinfo {year} {2023})}\BibitemShut {NoStop}%
\bibitem [{\citenamefont {Tarabunga}\ \emph {et~al.}(2022)\citenamefont {Tarabunga}, \citenamefont {Surace}, \citenamefont {Andreoni}, \citenamefont {Angelone},\ and\ \citenamefont {Dalmonte}}]{tarabunga:2022}%
  \BibitemOpen
  \bibfield  {author} {\bibinfo {author} {\bibfnamefont {P.~S.}\ \bibnamefont {Tarabunga}}, \bibinfo {author} {\bibfnamefont {F.~M.}\ \bibnamefont {Surace}}, \bibinfo {author} {\bibfnamefont {R.}~\bibnamefont {Andreoni}}, \bibinfo {author} {\bibfnamefont {A.}~\bibnamefont {Angelone}},\ and\ \bibinfo {author} {\bibfnamefont {M.}~\bibnamefont {Dalmonte}},\ }\bibfield  {title} {\bibinfo {title} {{Gauge-Theoretic Origin of Rydberg Quantum Spin Liquids}},\ }\href {https://doi.org/10.1103/PhysRevLett.129.195301} {\bibfield  {journal} {\bibinfo  {journal} {Phys. Rev. Lett.}\ }\textbf {\bibinfo {volume} {129}},\ \bibinfo {pages} {195301} (\bibinfo {year} {2022})}\BibitemShut {NoStop}%
\end{thebibliography}%

\clearpage
\onecolumngrid          
\appendix   
\begin{center}
  \large\textbf{Supplemental Material for}\\[6pt]
  \large\textit{``Incommensurate Spin-Density Waves in a Frustrated Maple-Leaf Lattice Ferromagnet''}\\[6pt]
  \normalsize Paul L. Ebert\,\orcidlink{0000-0003-1614-6920}, Yasir Iqbal\,\orcidlink{https://orcid.org/0000-0002-3387-0120}, Alexander Wietek\orcidlink{0000-0002-4367-3438}\\[6pt]
  \today
  
\end{center}
\setcounter{page}{1}
\setcounter{section}{0}
\setcounter{figure}{0}
\setcounter{table}{0}
\setcounter{equation}{0}
\renewcommand{\thefigure}{S\arabic{figure}}
\renewcommand{\thetable}{S\arabic{table}}
\renewcommand{\theequation}{S\arabic{equation}} 
\renewcommand{\thepage}{S-\arabic{page}} 

    \section{Two-Magnon Interactions above Singlet Ground State}

        Spin-nematic phases are typically understood in the context of multi-magnon interactions, where the interaction has to be sufficiently strong for collective multi-magnon excitations to be preferred over forming one or more non-interacting magnons~\cite{Shannon2006, momoi:2006}.
        Throughout the purple region in Fig.~\ref{fig:phase_diagram}, which we like to probe for signs of spin-nematic behavior, the ground state is a singlet $(S=0)$ such that the $n$-magnon sector has $S=n$.
        If we define $E_0(S)$ as the lowest energy in the spin-$S$ sector of the Hilbert space, we can give an estimate if, and to what degree, two-magnon interactions lower the excitation energy of the $S=2$ sector.
        The energy associated with a single 1-magnon excitation is $E^{\rm 1, free} = E_0(1) - E_0(0)$ and forming a 2-magnon state corresponds to $E^{\rm 2, interact.} = E_0( 2) - E_0(0)$.
        This means that
        \begin{equation}\label{eq:app:two-magnon} 
            \Delta_2 = E^{\rm 2, interact.} - 2 E^{\rm 1, free} =  E_0(2) - 2E_0(1) + E_0(0),
        \end{equation}
        becomes negative whenever two-magnon interactions lower the excitation energy of the $S=2$ level compared to what is expected for two non-interacting magnons.
        Note that a more accurate computation of single or multi-magnon excitations would be concerned with momentum conservation constraints, which we neglect here.
        To the best of our knowledge, it is a necessary condition for $d$-wave spin-nematic phases to have $\Delta_2 < 0$ and similar condition exist for $p$-wave spin-nematic regimes where three-magnon interactions become important~\cite{momoi:2006}.
        It is easy to see that $\Delta_2 < 0$ follows in clear-cut nematic phases where the $S=2$ level becomes the first excited state and is thus lower in energy than the $S=1$ level~\cite{wietek:2020}.

\begin{figure}[!h]
            \centering
            \includegraphics[width=0.45\linewidth]{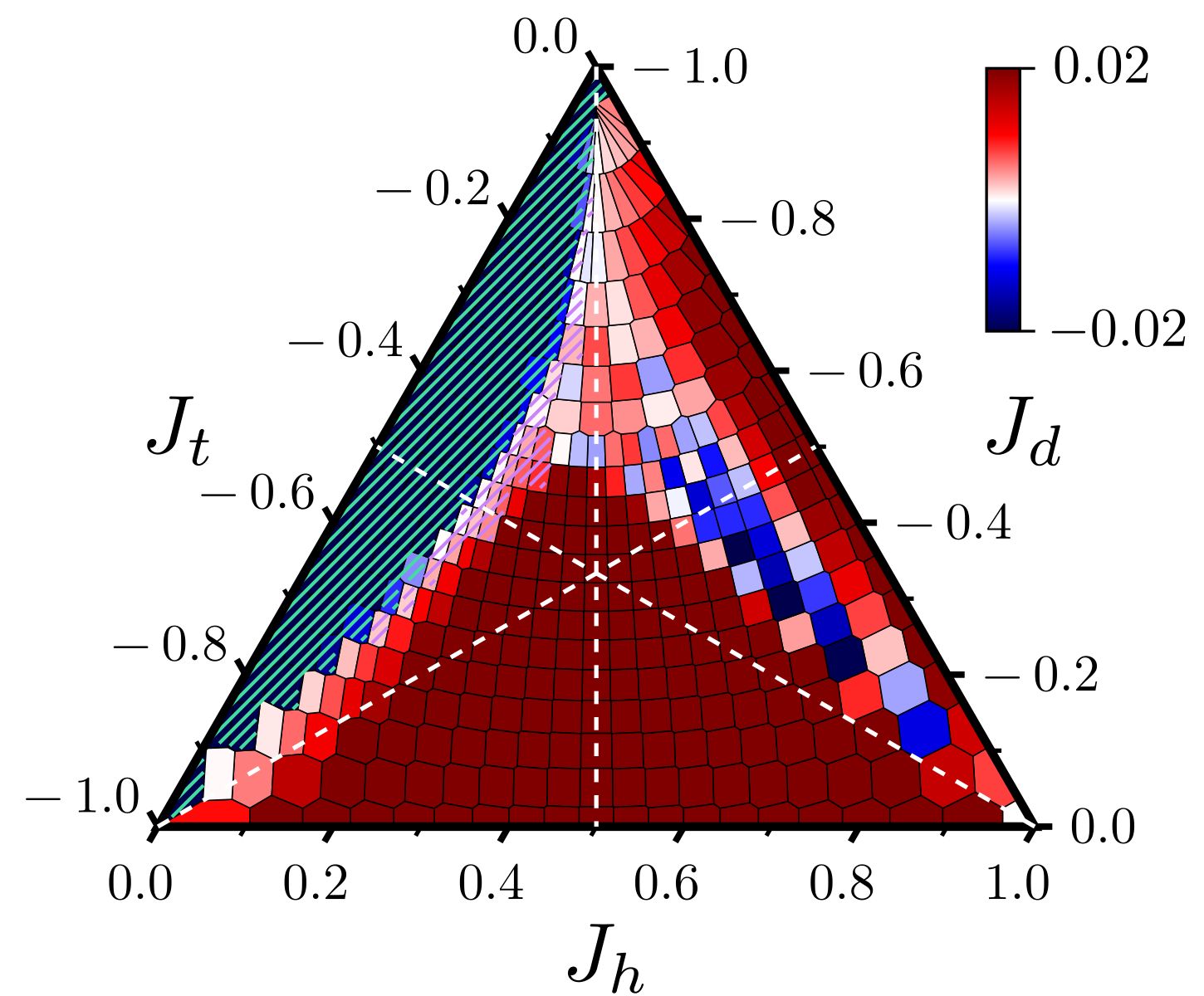}
            \caption{\label{fig:app:magnon}
            Two-magnon interaction energy $\Delta_2$ above a singlet ground state on the $N=36$ cluster, as defined in Eq.~\eqref{eq:app:two-magnon}, where negative (blue) values indicate that two-magnon interactions lower the energy of the $S=2$ level.
            Green (purple) lines are overlaid to show which data points are classified as the ferromagnetic (spin-density wave) phase in Fig.~\ref{fig:phase_diagram}.
            Note that the blue color in the ferromagnetic region is meaningless since here the ground state is not a singlet and hence $\Delta_2$ does not measure excitations above the ground state.
            }
        \end{figure}

        We show a heatmap of $\Delta_2$ in Fig.~\ref{fig:app:magnon} which is found to be zero or positive throughout the whole purple region near the ferromagnetic phase (note the purple and green lines respectively highlighting the extent of both phases). We therefore find no evidence for a $d$-wave spin-nematic phase in the purple region where spin-spin correlations display wave patterns.

    \section{Ground State Energies as a Function of $S$ near the Ferromagnetic Phase}

        As mentioned earlier, our ED data does not feature low-lying $S=2$ or $S=3$ levels that could lead to a zero-field spin-nematic phase.
        Moreover, no obvious anomalies are observed in the high-$S$ regions of the spectrum which would warrant the expectation of a spin-nematic phase in non-zero magnetic fields (as e.g. in Ref.~\cite{Shannon2006}).
        To illustrate this, we here show the full spectra at the two points corresponding to the insets in Fig.~\ref{fig:N-36-N1-N4-cut}.
        The exact coordinates for subplot (a) are $J_t = -0.45$,  $J_h = 0.146$,  $J_d = -0.404$ or $\theta/\pi = 0.275$, $\phi/\pi = 0.1$ and for subplot (b) $J_t = -0.145$, $J_h = 0.106$, $J_d = -0.749$ or $\theta/\pi = 0.075$, $\phi/\pi = 0.2$, where the $\theta$ and $\phi$ angles refer to the spherical parametrization introduced in Sec. D below (see Eq.~\eqref{eq:sphericalParametrization}).
        Both plots only show the lowest-energy level for each space-group irrep on the $N=36$ cluster in each spin-$S$ sector.

        The curve $E_0(S)$ (the lowest energy level with SU$(2)$ quantum number $S$) that is observed throughout nearly all points in the spin density wave regime near the FM boundary is exemplified in (a).
        It is monotonically increasing (in particular no $S=2$ or $S=3$ level becomes the first excited state) and its low-$S$ shape is generally convex (indicating e.g. a repulsive two-magnon interaction via Eq.~\eqref{eq:app:two-magnon}).
        The high-$S$ levels are observed to be only slightly concave and in most cases close to linear in $S$ which implies that the prospects of observing a spin-nematic phase at non-zero fields in which these levels become the first excited states are low.

        As briefly mentioned in the main text, our ED scans revealed a single point in the upper part of the phase diagram in Fig.~\ref{fig:phase_diagram}{a} at which $E_0(S)$ follows an arch-type curve.
        The spectrum of this point, most likely astonishingly close to the FM phase transition, is shown in (b).
        Here the ground state is still in the non-polarized $S=0$ sector while the fully-polarized $S=S_{\rm max}$ level becomes the sixth excited state (also see the inset in Fig.~\ref{fig:N-36-N1-N4-cut}{b}).
        The system can thus be polarized by a small excitation and we generally expect $E_0(S)$ to maintain its shape moving closer towards the FM boundary, at which point the ground state jumps from $S=0$ to $S=S_{\rm max}$.
        This, given that the shape of $E_0(S)$ is retained as system size is increased, presents a strong argument for a first-order transition in the $J_t / J_h \lesssim 0, J_d / J_h \ll 0$ regime of the FM phase boundary during which the magnetization jumps abruptly.
        We are currently unaware of another published ED spectrum so close to a FM phase boundary that $E_0(S)$ displays an arch-shape.

        \begin{figure}[!h]
            \centering
            \includegraphics[width=0.8\linewidth]{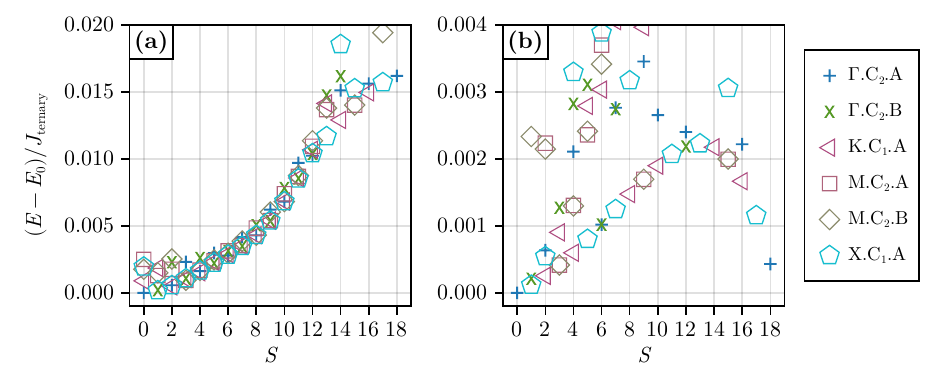}
            \caption{\label{fig:app:tos} Tower-of-states plot on the $N=36$ cluster at the two exemplary spin-density-wave points shown as insets in Fig.~\ref{fig:N-36-N1-N4-cut} (a) and (b), respectively.
            The energy scale was chosen as to enforce $1 = J_{\rm ternary} = |J_t| + J_h + |J_d|$.
            For each $S$ we show the single lowest energy level in each space group irrep (see legend on the right).
            Note that only (a) displays the expected monotonous behavior while (b) shows an arch-shaped dependence of the ground state energy in each $S$ sector.
            }
        \end{figure}

    \section{Finite Size Clusters}

\begin{figure}[!h]
            \centering
            \includegraphics[width=0.8\linewidth]{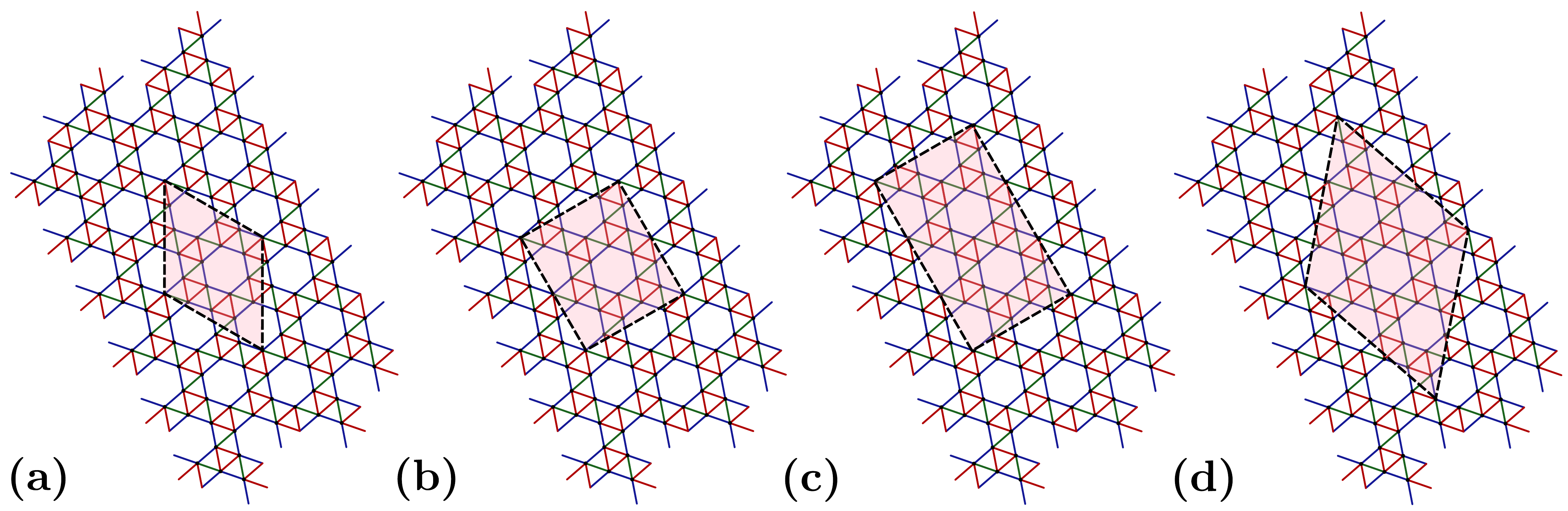}
            \caption{\label{fig:app:cluster}
            Different finite size cluster used throughout this study:
            (a) $N=18$, (b) $N=24$, (c) $N=36$, (d) $N=42$.
            }
        \end{figure}

        In total, four different finite-size clusters of the MLL were used throughout this work, which are shown in Fig.~\ref{fig:app:cluster}.
        For each of the four system sizes $N \in \{42, 36, 24, 18\}$ we picked the most symmetric cluster of that size in the sense that important high-symmetry momenta are resolved (e.g. K$_{\rm MLL}$, M$_{\rm MLL}$, or X$_{\rm MLL}$ as defined in Fig.~\ref{fig:structureFactor}; the MLL subscript being dropped from now on).

        Adopting the notation $\bm{k}.\hat{\mathcal{G}}(\bm{k})$ to denote that a cluster resolves momentum $\bm{k}$ and that the point group of the associated little group of $\bm{k}$ on that cluster is $\hat{\mathcal{G}}(\bm{k})$, the symmetries of the employed clusters can be summarized as follows:
        For $N=18$ we resolve $\Gamma$.C$_6$ and K.C$_3$ (i.e. the full symmetries of the infinite MLL at both momenta).
        For $N=24$ we resolve $\Gamma$.C$_2$, M.C$_2$, and Y.C$_1$ where Y lies on the $\Gamma$-K line between X and K.
        For $N=36$ we resolve $\Gamma$.C$_2$, K.C$_1$, M.C$_2$, and X.C$_1$, which makes it the most symmetric currently accessible cluster for phase diagram scans.
        For $N=42$ we resolve $\Gamma$.C$_6$ and O.C$_1$, where O lies inside the Brillouin zone roughly between M and X.
        
    \section{Table of Classified Points and Couplings}
    \setcounter{equation}{1}

    The phase diagram of the Hamiltonian~\eqref{eq:H} with arbitrary $J_t, J_h, J_d \in \mathbb{R}$ can be mapped to the $J_t^2 + J_h^2 + J_d^2 = 1$ sphere by appropriately fixing the overall energy scale.
    Our scans of the $J_t, J_d \leq 0$ (ferromagnetic) $0 \leq J_h$ (antiferromagnetic) octant of this sphere are based on the spherical parametrization
    \begin{align}\label{eq:sphericalParametrization}
        &J_t = -\cos\phi\sin\theta,&
        &J_h = +\sin\phi\sin\theta,&
        &J_d = -\cos\theta,&
    \end{align}
    under which the octant of interest is parametrized by $\phi, \theta \in [0, \pi/2]$.
    In the following, we show all coupling values for which at least the $S^z_{\rm tot} = \sum_i S^z_i = 0, 1, 2$ sectors were scanned.
    At some data points we scanned considerably more sectors (see e.g. Fig.~\ref{fig:N-36-N1-N4-cut}) and sometimes even every sector (see e.g. Fig.~\ref{fig:app:tos}).

\setlength{\LTcapwidth}{\textwidth}
\begin{longtable*}{|r|l|l|c|c|c|c|l|}
\caption{\label{tab:classification} List of coupling values inside the phase diagram scan of the Hamiltonian~\eqref{eq:H} with $J_t, J_d \leq 0 \leq J_h$.
The angles $\theta, \phi \in [0, \pi/2]$ describe this model under the spherical parametrization from Eq.~\eqref{eq:sphericalParametrization} (i.e. under the $J_t^2 + J_h^2 + J_d^2 = 1$ constraint).
The values given for $J_t, J_h, J_d$ assume the ternary normalization $1 = |J_t| + J_h + |J_d|$ that we used throughout the main text and employed in the phase diagram in Fig.~\ref{fig:phase_diagram}{a}.
The ground state energies (per spin) also assume the ternary normalization and were obtained on the $N=36$ cluster.
}  \\
\hline
\hline
Index & $\theta/\pi$ & $\phi/\pi$ & $J_t$ & $J_h$ & $J_d$ & classified phase & $E_0/N/J_{\rm ternary}$  \\
\hline
\endfirsthead 

\multicolumn{8}{c}{{\tablename\ \thetable{} -- continued from previous page}} \\
\hline
Index & $\theta/\pi$ & $\phi/\pi$ & $J_t$ & $J_h$ & $J_d$ & classified phase & $E_0/N/J_{\rm ternary}$ \\
\hline
\endhead 

\hline
\multicolumn{8}{r}{{Continued on next page}} \\
\endfoot 

\hline
\hline
\endlastfoot 

1 & 0 & 0 & 0 & 0 & -1 & isolated FM $J_d$ - ``dimers'' & -0.125 \\
2 & 0.025 & 0 & -0.07296 & 0 & -0.92704 & FM & -0.13411996 \\
3 & 0.025 & 0.05 & -0.071312 & 0.011295 & -0.91739 & FM & -0.12967845 \\
4 & 0.025 & 0.1 & -0.068097 & 0.022126 & -0.90978 & FM & -0.12521485 \\
5 & 0.025 & 0.15 & -0.063411 & 0.03231 & -0.90428 & FM & -0.12081028 \\
6 & 0.025 & 0.2 & -0.057365 & 0.041678 & -0.90096 & FM & -0.11654135 \\
7 & 0.025 & 0.225 & -0.053868 & 0.046008 & -0.90012 & FM & -0.11448061 \\
8 & 0.025 & 0.25 & -0.050077 & 0.050077 & -0.89985 & inconclusive & -0.11411144 \\
9 & 0.025 & 0.3 & -0.041678 & 0.057365 & -0.90096 & inconclusive & -0.11723839 \\
10 & 0.025 & 0.35 & -0.03231 & 0.063411 & -0.90428 & inconclusive & -0.12061396 \\
11 & 0.025 & 0.4 & -0.022126 & 0.068097 & -0.90978 & inconclusive & -0.1241199 \\
12 & 0.025 & 0.45 & -0.011295 & 0.071312 & -0.91739 & inconclusive & -0.12767863 \\
13 & 0.025 & 0.5 & 0 & 0.07296 & -0.92704 & inconclusive & -0.13121918 \\
14 & 0.05 & 0 & -0.13673 & 0 & -0.86327 & FM & -0.14209109 \\
15 & 0.05 & 0.05 & -0.13244 & 0.020976 & -0.84659 & FM & -0.13368855 \\
16 & 0.05 & 0.1 & -0.12557 & 0.040801 & -0.83363 & FM & -0.12539619 \\
17 & 0.05 & 0.15 & -0.11634 & 0.059277 & -0.82438 & FM & -0.11731328 \\
18 & 0.05 & 0.2 & -0.10492 & 0.076231 & -0.81885 & FM & -0.10952869 \\
19 & 0.05 & 0.225 & -0.098452 & 0.084086 & -0.81746 & spin density wave & -0.10640483 \\
20 & 0.05 & 0.25 & -0.0915 & 0.0915 & -0.817 & inconclusive & -0.10820926 \\
21 & 0.05 & 0.3 & -0.076231 & 0.10492 & -0.81885 & canted $120^{\circ}$ & -0.11426367 \\
22 & 0.05 & 0.35 & -0.059277 & 0.11634 & -0.82438 & canted $120^{\circ}$ & -0.12069881 \\
23 & 0.05 & 0.4 & -0.040801 & 0.12557 & -0.83363 & canted $120^{\circ}$ & -0.12731359 \\
24 & 0.05 & 0.45 & -0.020976 & 0.13244 & -0.84659 & canted $120^{\circ}$ & -0.13398745 \\
25 & 0.05 & 0.5 & 0 & 0.13673 & -0.86327 & canted $120^{\circ}$ & -0.14062242 \\
26 & 0.075 & 0 & -0.1936 & 0 & -0.8064 & FM & -0.14919995 \\
27 & 0.075 & 0.05 & -0.18603 & 0.029464 & -0.78451 & FM & -0.13720436 \\
28 & 0.075 & 0.1 & -0.1753 & 0.056958 & -0.76774 & FM & -0.12555308 \\
29 & 0.075 & 0.15 & -0.1617 & 0.082389 & -0.75591 & FM & -0.11431625 \\
30 & 0.075 & 0.1625 & -0.15788 & 0.088416 & -0.75371 & FM & -0.11157883 \\
31 & 0.075 & 0.175 & -0.15389 & 0.094306 & -0.7518 & FM & -0.10887186 \\
32 & 0.075 & 0.2 & -0.14545 & 0.10568 & -0.74887 & spin density wave & -0.1035646 \\
33 & 0.075 & 0.225 & -0.13639 & 0.11649 & -0.74712 & inconclusive & -0.10224623 \\
34 & 0.075 & 0.25 & -0.12673 & 0.12673 & -0.74653 & inconclusive & -0.10612757 \\
35 & 0.075 & 0.3 & -0.10568 & 0.14545 & -0.74887 & canted $120^{\circ}$ & -0.11485434 \\
36 & 0.075 & 0.35 & -0.082389 & 0.1617 & -0.75591 & canted $120^{\circ}$ & -0.12400565 \\
37 & 0.075 & 0.4 & -0.056958 & 0.1753 & -0.76774 & canted $120^{\circ}$ & -0.13334582 \\
38 & 0.075 & 0.45 & -0.029464 & 0.18603 & -0.78451 & canted $120^{\circ}$ & -0.14275356 \\
39 & 0.075 & 0.5 & 0 & 0.1936 & -0.8064 & canted $120^{\circ}$ & -0.15214618 \\
40 & 0.1 & 0 & -0.24524 & 0 & -0.75476 & FM & -0.15565466 \\
41 & 0.1 & 0.05 & -0.23395 & 0.037054 & -0.729 & FM & -0.14034843 \\
42 & 0.1 & 0.1 & -0.21925 & 0.071239 & -0.70951 & FM & -0.12569176 \\
43 & 0.1 & 0.15 & -0.20146 & 0.10265 & -0.69589 & FM & -0.11168893 \\
44 & 0.1 & 0.1625 & -0.19656 & 0.11008 & -0.69336 & FM & -0.10829031 \\
45 & 0.1 & 0.175 & -0.19148 & 0.11734 & -0.69118 & FM & -0.10493252 \\
46 & 0.1 & 0.2 & -0.18081 & 0.13136 & -0.68783 & spin density wave & -0.10032425 \\
47 & 0.1 & 0.25 & -0.15742 & 0.15742 & -0.68516 & inconclusive & -0.10699477 \\
48 & 0.1 & 0.3 & -0.13136 & 0.18081 & -0.68783 & canted $120^{\circ}$ & -0.11803905 \\
49 & 0.1 & 0.35 & -0.10265 & 0.20146 & -0.69589 & canted $120^{\circ}$ & -0.12951705 \\
50 & 0.1 & 0.4 & -0.071239 & 0.21925 & -0.70951 & canted $120^{\circ}$ & -0.14120423 \\
51 & 0.1 & 0.45 & -0.037054 & 0.23395 & -0.729 & canted $120^{\circ}$ & -0.15301534 \\
52 & 0.1 & 0.5 & 0 & 0.24524 & -0.75476 & canted $120^{\circ}$ & -0.16491625 \\
53 & 0.125 & 0 & -0.29289 & 0 & -0.70711 & FM & -0.16161165 \\
54 & 0.125 & 0.05 & -0.27757 & 0.043963 & -0.67847 & FM & -0.14321023 \\
55 & 0.125 & 0.1 & -0.25884 & 0.084103 & -0.65706 & FM & -0.12581667 \\
56 & 0.125 & 0.15 & -0.23702 & 0.12077 & -0.64221 & FM & -0.10933963 \\
57 & 0.125 & 0.1625 & -0.2311 & 0.12942 & -0.63947 & FM & -0.10535378 \\
58 & 0.125 & 0.175 & -0.22501 & 0.13789 & -0.6371 & spin density wave & -0.10180365 \\
59 & 0.125 & 0.2 & -0.21228 & 0.15423 & -0.63348 & spin density wave & -0.099883595 \\
60 & 0.125 & 0.25 & -0.1847 & 0.1847 & -0.6306 & inconclusive & -0.11017238 \\
61 & 0.125 & 0.3 & -0.15423 & 0.21228 & -0.63348 & inconclusive & -0.12311615 \\
62 & 0.125 & 0.35 & -0.12077 & 0.23702 & -0.64221 & canted $120^{\circ}$ & -0.13652068 \\
63 & 0.125 & 0.4 & -0.084103 & 0.25884 & -0.65706 & canted $120^{\circ}$ & -0.1502066 \\
64 & 0.125 & 0.45 & -0.043963 & 0.27757 & -0.67847 & canted $120^{\circ}$ & -0.16414778 \\
65 & 0.125 & 0.5 & 0 & 0.29289 & -0.70711 & canted $120^{\circ}$ & -0.17838092 \\
66 & 0.15 & 0 & -0.33754 & 0 & -0.66246 & FM & -0.16719252 \\
67 & 0.15 & 0.05 & -0.31792 & 0.050353 & -0.63173 & FM & -0.14585732 \\
68 & 0.15 & 0.1 & -0.29511 & 0.095888 & -0.609 & FM & -0.12593112 \\
69 & 0.15 & 0.15 & -0.26938 & 0.13726 & -0.59336 & FM & -0.10720145 \\
70 & 0.15 & 0.1625 & -0.26251 & 0.14701 & -0.59048 & spin density wave & -0.10324734 \\
71 & 0.15 & 0.175 & -0.25545 & 0.15654 & -0.588 & spin density wave & -0.10173684 \\
72 & 0.15 & 0.2 & -0.24082 & 0.17497 & -0.58421 & inconclusive & -0.10235917 \\
73 & 0.15 & 0.25 & -0.2094 & 0.2094 & -0.5812 & inconclusive & -0.11522796 \\
74 & 0.15 & 0.3 & -0.17497 & 0.24082 & -0.58421 & inconclusive & -0.12961545 \\
75 & 0.15 & 0.35 & -0.13726 & 0.26938 & -0.59336 & inconclusive & -0.14456108 \\
76 & 0.15 & 0.4 & -0.095888 & 0.29511 & -0.609 & canted $120^{\circ}$ & -0.15993983 \\
77 & 0.15 & 0.45 & -0.050353 & 0.31792 & -0.63173 & canted $120^{\circ}$ & -0.17579256 \\
78 & 0.15 & 0.5 & 0 & 0.33754 & -0.66246 & canted $120^{\circ}$ & -0.1922403 \\
79 & 0.175 & 0 & -0.37996 & 0 & -0.62004 & FM & -0.17249508 \\
80 & 0.175 & 0.05 & -0.3558 & 0.056353 & -0.58785 & FM & -0.14834248 \\
81 & 0.175 & 0.1 & -0.32887 & 0.10686 & -0.56428 & FM & -0.12603761 \\
82 & 0.175 & 0.125 & -0.31441 & 0.13023 & -0.55535 & FM & -0.11546379 \\
83 & 0.175 & 0.1375 & -0.30694 & 0.1415 & -0.55155 & FM & -0.1103043 \\
84 & 0.175 & 0.15 & -0.29931 & 0.15251 & -0.54818 & spin density wave & -0.10598752 \\
85 & 0.175 & 0.2 & -0.26712 & 0.19407 & -0.5388 & inconclusive & -0.10801606 \\
86 & 0.175 & 0.25 & -0.23214 & 0.23214 & -0.53572 & hexagonal singlet & -0.12184519 \\
87 & 0.175 & 0.3 & -0.19407 & 0.26712 & -0.5388 & hexagonal singlet & -0.13722382 \\
88 & 0.175 & 0.35 & -0.15251 & 0.29931 & -0.54818 & inconclusive & -0.15335999 \\
89 & 0.175 & 0.4 & -0.10686 & 0.32887 & -0.56428 & canted $120^{\circ}$ & -0.17017244 \\
90 & 0.175 & 0.45 & -0.056353 & 0.3558 & -0.58785 & canted $120^{\circ}$ & -0.18776706 \\
91 & 0.175 & 0.5 & 0 & 0.37996 & -0.62004 & canted $120^{\circ}$ & -0.20635884 \\
92 & 0.2 & 0 & -0.42081 & 0 & -0.57919 & FM & -0.17760097 \\
93 & 0.2 & 0.025 & -0.40661 & 0.032001 & -0.56139 & FM & -0.16382622 \\
94 & 0.2 & 0.05 & -0.39186 & 0.062065 & -0.54607 & FM & -0.15070839 \\
95 & 0.2 & 0.075 & -0.37657 & 0.090406 & -0.53303 & FM & -0.13816872 \\
96 & 0.2 & 0.1 & -0.36073 & 0.11721 & -0.52206 & FM & -0.12613815 \\
97 & 0.2 & 0.125 & -0.34435 & 0.14264 & -0.51301 & FM & -0.11455575 \\
98 & 0.2 & 0.1375 & -0.33595 & 0.15488 & -0.50917 & spin density wave & -0.10980965 \\
99 & 0.2 & 0.15 & -0.32741 & 0.16682 & -0.50577 & spin density wave & -0.10826538 \\
100 & 0.2 & 0.175 & -0.30988 & 0.18989 & -0.50023 & spin density wave & -0.10987365 \\
101 & 0.2 & 0.2 & -0.29173 & 0.21195 & -0.49632 & inconclusive & -0.11542369 \\
102 & 0.2 & 0.225 & -0.27292 & 0.23309 & -0.49399 & inconclusive & -0.12231738 \\
103 & 0.2 & 0.25 & -0.25339 & 0.25339 & -0.49322 & hexagonal singlet & -0.12976895 \\
104 & 0.2 & 0.275 & -0.23309 & 0.27292 & -0.49399 & hexagonal singlet & -0.13760019 \\
105 & 0.2 & 0.3 & -0.21195 & 0.29173 & -0.49632 & hexagonal singlet & -0.14573034 \\
106 & 0.2 & 0.325 & -0.18989 & 0.30988 & -0.50023 & hexagonal singlet & -0.15411987 \\
107 & 0.2 & 0.35 & -0.16682 & 0.32741 & -0.50577 & hexagonal singlet & -0.16275499 \\
108 & 0.2 & 0.375 & -0.14264 & 0.34435 & -0.51301 & inconclusive & -0.17163931 \\
109 & 0.2 & 0.4 & -0.11721 & 0.36073 & -0.52206 & canted $120^{\circ}$ & -0.180789 \\
110 & 0.2 & 0.425 & -0.090406 & 0.37657 & -0.53303 & canted $120^{\circ}$ & -0.19023032 \\
111 & 0.2 & 0.45 & -0.062065 & 0.39186 & -0.54607 & canted $120^{\circ}$ & -0.19999863 \\
112 & 0.2 & 0.475 & -0.032001 & 0.40661 & -0.56139 & canted $120^{\circ}$ & -0.21013833 \\
113 & 0.2 & 0.5 & 0 & 0.42081 & -0.57919 & canted $120^{\circ}$ & -0.22070351 \\
114 & 0.225 & 0 & -0.46065 & 0 & -0.53935 & FM & -0.18258114 \\
115 & 0.225 & 0.025 & -0.44382 & 0.034929 & -0.52125 & FM & -0.16737886 \\
116 & 0.225 & 0.05 & -0.42665 & 0.067575 & -0.50577 & FM & -0.15299088 \\
117 & 0.225 & 0.075 & -0.40913 & 0.098224 & -0.49264 & FM & -0.13930754 \\
118 & 0.225 & 0.1 & -0.39123 & 0.12712 & -0.48165 & FM & -0.12623438 \\
119 & 0.225 & 0.1125 & -0.38213 & 0.14098 & -0.47689 & FM & -0.11990054 \\
120 & 0.225 & 0.125 & -0.37292 & 0.15447 & -0.47261 & spin density wave & -0.11457716 \\
121 & 0.225 & 0.1375 & -0.3636 & 0.16762 & -0.46878 & spin density wave & -0.11301951 \\
122 & 0.225 & 0.15 & -0.35416 & 0.18045 & -0.46539 & spin density wave & -0.11358425 \\
123 & 0.225 & 0.175 & -0.3349 & 0.20522 & -0.45988 & spin density wave & -0.11809405 \\
124 & 0.225 & 0.2 & -0.31508 & 0.22892 & -0.456 & inconclusive & -0.12447076 \\
125 & 0.225 & 0.225 & -0.29465 & 0.25166 & -0.45369 & inconclusive & -0.13141215 \\
126 & 0.225 & 0.25 & -0.27354 & 0.27354 & -0.45293 & inconclusive & -0.13883877 \\
127 & 0.225 & 0.275 & -0.25166 & 0.29465 & -0.45369 & hexagonal singlet & -0.14672038 \\
128 & 0.225 & 0.3 & -0.22892 & 0.31508 & -0.456 & hexagonal singlet & -0.15500764 \\
129 & 0.225 & 0.325 & -0.20522 & 0.3349 & -0.45988 & hexagonal singlet & -0.16366106 \\
130 & 0.225 & 0.35 & -0.18045 & 0.35416 & -0.46539 & hexagonal singlet & -0.17266366 \\
131 & 0.225 & 0.375 & -0.15447 & 0.37292 & -0.47261 & inconclusive & -0.18201949 \\
132 & 0.225 & 0.4 & -0.12712 & 0.39123 & -0.48165 & inconclusive & -0.19174945 \\
133 & 0.225 & 0.425 & -0.098224 & 0.40913 & -0.49264 & canted $120^{\circ}$ & -0.20188847 \\
134 & 0.225 & 0.45 & -0.067575 & 0.42665 & -0.50577 & canted $120^{\circ}$ & -0.21248443 \\
135 & 0.225 & 0.475 & -0.034929 & 0.44382 & -0.52125 & canted $120^{\circ}$ & -0.22359851 \\
136 & 0.225 & 0.5 & 0 & 0.46065 & -0.53935 & canted $120^{\circ}$ & -0.23530675 \\
137 & 0.25 & 0 & -0.5 & 0 & -0.5 & FM & -0.1875 \\
138 & 0.25 & 0.025 & -0.48035 & 0.037805 & -0.48184 & FM & -0.17086759 \\
139 & 0.25 & 0.05 & -0.46065 & 0.07296 & -0.46639 & FM & -0.15522127 \\
140 & 0.25 & 0.075 & -0.44082 & 0.10583 & -0.45335 & FM & -0.14041572 \\
141 & 0.25 & 0.1 & -0.42081 & 0.13673 & -0.44246 & FM & -0.1263277 \\
142 & 0.25 & 0.1125 & -0.41071 & 0.15152 & -0.43777 & spin density wave & -0.12024641 \\
143 & 0.25 & 0.125 & -0.40054 & 0.16591 & -0.43355 & spin density wave & -0.11868162 \\
144 & 0.25 & 0.1375 & -0.3903 & 0.17993 & -0.42977 & spin density wave & -0.11983474 \\
145 & 0.25 & 0.15 & -0.37996 & 0.1936 & -0.42644 & inconclusive & -0.12214282 \\
146 & 0.25 & 0.175 & -0.35899 & 0.21999 & -0.42103 & collinear N\'eel AFM & -0.12833451 \\
147 & 0.25 & 0.2 & -0.33754 & 0.24524 & -0.41722 & collinear N\'eel AFM & -0.13487654 \\
148 & 0.25 & 0.225 & -0.31554 & 0.2695 & -0.41496 & collinear N\'eel AFM & -0.14171285 \\
149 & 0.25 & 0.25 & -0.29289 & 0.29289 & -0.41421 & inconclusive & -0.14896071 \\
150 & 0.25 & 0.275 & -0.2695 & 0.31554 & -0.41496 & inconclusive & -0.15671781 \\
151 & 0.25 & 0.3 & -0.24524 & 0.33754 & -0.41722 & hexagonal singlet & -0.16500441 \\
152 & 0.25 & 0.325 & -0.21999 & 0.35899 & -0.42103 & hexagonal singlet & -0.17379423 \\
153 & 0.25 & 0.35 & -0.1936 & 0.37996 & -0.42644 & hexagonal singlet & -0.18306413 \\
154 & 0.25 & 0.375 & -0.16591 & 0.40054 & -0.43355 & inconclusive & -0.19281368 \\
155 & 0.25 & 0.4 & -0.13673 & 0.42081 & -0.44246 & inconclusive & -0.20306579 \\
156 & 0.25 & 0.425 & -0.10583 & 0.44082 & -0.45335 & inconclusive & -0.21386356 \\
157 & 0.25 & 0.45 & -0.07296 & 0.46065 & -0.46639 & canted $120^{\circ}$ & -0.2252687 \\
158 & 0.25 & 0.475 & -0.037805 & 0.48035 & -0.48184 & canted $120^{\circ}$ & -0.23736207 \\
159 & 0.25 & 0.5 & 0 & 0.5 & -0.5 & canted $120^{\circ}$ & -0.25024623 \\
160 & 0.275 & 0 & -0.53935 & 0 & -0.46065 & FM & -0.19241886 \\
161 & 0.275 & 0.025 & -0.51668 & 0.040664 & -0.44265 & FM & -0.17433642 \\
162 & 0.275 & 0.05 & -0.49429 & 0.078288 & -0.42742 & FM & -0.15742819 \\
163 & 0.275 & 0.075 & -0.47205 & 0.11333 & -0.41462 & FM & -0.14150774 \\
164 & 0.275 & 0.0875 & -0.46095 & 0.13 & -0.40905 & FM & -0.13386827 \\
165 & 0.275 & 0.1 & -0.44985 & 0.14617 & -0.40398 & spin density wave & -0.12686931 \\
166 & 0.275 & 0.1125 & -0.43874 & 0.16186 & -0.3994 & spin density wave & -0.12532098 \\
167 & 0.275 & 0.125 & -0.42759 & 0.17712 & -0.39529 & spin density wave & -0.12695078 \\
168 & 0.275 & 0.15 & -0.40517 & 0.20645 & -0.38838 & collinear N\'eel AFM & -0.13309923 \\
169 & 0.275 & 0.175 & -0.38248 & 0.23439 & -0.38313 & collinear N\'eel AFM & -0.13960434 \\
170 & 0.275 & 0.2 & -0.35942 & 0.26114 & -0.37944 & collinear N\'eel AFM & -0.14619782 \\
171 & 0.275 & 0.225 & -0.33588 & 0.28687 & -0.37726 & collinear N\'eel AFM & -0.15295323 \\
172 & 0.275 & 0.25 & -0.31174 & 0.31174 & -0.37653 & collinear N\'eel AFM & -0.16001763 \\
173 & 0.275 & 0.275 & -0.28687 & 0.33588 & -0.37726 & inconclusive & -0.16756679 \\
174 & 0.275 & 0.3 & -0.26114 & 0.35942 & -0.37944 & inconclusive & -0.17572803 \\
175 & 0.275 & 0.325 & -0.23439 & 0.38248 & -0.38313 & hexagonal singlet & -0.18453771 \\
176 & 0.275 & 0.35 & -0.20645 & 0.40517 & -0.38838 & hexagonal singlet & -0.19398373 \\
177 & 0.275 & 0.375 & -0.17712 & 0.42759 & -0.39529 & hexagonal singlet & -0.20406042 \\
178 & 0.275 & 0.4 & -0.14617 & 0.44985 & -0.40398 & hexagonal singlet & -0.21478937 \\
179 & 0.275 & 0.425 & -0.11333 & 0.47205 & -0.41462 & inconclusive & -0.22622065 \\
180 & 0.275 & 0.45 & -0.078288 & 0.49429 & -0.42742 & canted $120^{\circ}$ & -0.23843101 \\
181 & 0.275 & 0.475 & -0.040664 & 0.51668 & -0.44265 & canted $120^{\circ}$ & -0.25152434 \\
182 & 0.275 & 0.5 & 0 & 0.53935 & -0.46065 & canted $120^{\circ}$ & -0.26563529 \\
183 & 0.3 & 0 & -0.57919 & 0 & -0.42081 & FM & -0.19739903 \\
184 & 0.3 & 0.025 & -0.55325 & 0.043542 & -0.4032 & FM & -0.17782841 \\
185 & 0.3 & 0.05 & -0.52799 & 0.083625 & -0.38839 & FM & -0.15963908 \\
186 & 0.3 & 0.075 & -0.5032 & 0.12081 & -0.37599 & FM & -0.14259727 \\
187 & 0.3 & 0.0875 & -0.49094 & 0.13846 & -0.3706 & spin density wave & -0.13466549 \\
188 & 0.3 & 0.1 & -0.47873 & 0.15555 & -0.36572 & spin density wave & -0.13283661 \\
189 & 0.3 & 0.1125 & -0.46656 & 0.17212 & -0.36131 & collinear N\'eel AFM & -0.13507543 \\
190 & 0.3 & 0.125 & -0.45442 & 0.18823 & -0.35736 & collinear N\'eel AFM & -0.13832926 \\
191 & 0.3 & 0.15 & -0.43012 & 0.21916 & -0.35073 & collinear N\'eel AFM & -0.14497304 \\
192 & 0.3 & 0.175 & -0.40569 & 0.24861 & -0.3457 & collinear N\'eel AFM & -0.15160197 \\
193 & 0.3 & 0.2 & -0.38101 & 0.27682 & -0.34217 & collinear N\'eel AFM & -0.15823158 \\
194 & 0.3 & 0.225 & -0.35593 & 0.30399 & -0.34008 & collinear N\'eel AFM & -0.16494577 \\
195 & 0.3 & 0.25 & -0.33031 & 0.33031 & -0.33939 & collinear N\'eel AFM & -0.17188426 \\
196 & 0.3 & 0.275 & -0.30399 & 0.35593 & -0.34008 & inconclusive & -0.17923984 \\
197 & 0.3 & 0.3 & -0.27682 & 0.38101 & -0.34217 & inconclusive & -0.18721726 \\
198 & 0.3 & 0.325 & -0.24861 & 0.40569 & -0.3457 & hexagonal singlet & -0.19595374 \\
199 & 0.3 & 0.35 & -0.21916 & 0.43012 & -0.35073 & hexagonal singlet & -0.20549186 \\
200 & 0.3 & 0.375 & -0.18823 & 0.45442 & -0.35736 & hexagonal singlet & -0.21583554 \\
201 & 0.3 & 0.4 & -0.15555 & 0.47873 & -0.36572 & hexagonal singlet & -0.22700514 \\
202 & 0.3 & 0.425 & -0.12081 & 0.5032 & -0.37599 & inconclusive & -0.23905624 \\
203 & 0.3 & 0.45 & -0.083625 & 0.52799 & -0.38839 & inconclusive & -0.25208157 \\
204 & 0.3 & 0.475 & -0.043542 & 0.55325 & -0.4032 & canted $120^{\circ}$ & -0.26621181 \\
205 & 0.3 & 0.5 & 0 & 0.57919 & -0.42081 & canted $120^{\circ}$ & -0.28162043 \\
206 & 0.325 & 0 & -0.62004 & 0 & -0.37996 & FM & -0.20250492 \\
207 & 0.325 & 0.025 & -0.59053 & 0.046476 & -0.363 & FM & -0.18138774 \\
208 & 0.325 & 0.05 & -0.56217 & 0.089039 & -0.34879 & FM & -0.16188158 \\
209 & 0.325 & 0.075 & -0.53468 & 0.12836 & -0.33696 & FM & -0.14369786 \\
210 & 0.325 & 0.0875 & -0.52118 & 0.14699 & -0.33184 & spin density wave & -0.14118839 \\
211 & 0.325 & 0.1 & -0.50781 & 0.165 & -0.3272 & collinear N\'eel AFM & -0.14398527 \\
212 & 0.325 & 0.125 & -0.48135 & 0.19938 & -0.31927 & collinear N\'eel AFM & -0.1508139 \\
213 & 0.325 & 0.15 & -0.45511 & 0.23189 & -0.31301 & collinear N\'eel AFM & -0.15759222 \\
214 & 0.325 & 0.175 & -0.42891 & 0.26283 & -0.30826 & collinear N\'eel AFM & -0.16428449 \\
215 & 0.325 & 0.2 & -0.40258 & 0.29249 & -0.30494 & collinear N\'eel AFM & -0.17093622 \\
216 & 0.325 & 0.225 & -0.37594 & 0.32109 & -0.30297 & collinear N\'eel AFM & -0.17763137 \\
217 & 0.325 & 0.25 & -0.34884 & 0.34884 & -0.30232 & collinear N\'eel AFM & -0.18449751 \\
218 & 0.325 & 0.275 & -0.32109 & 0.37594 & -0.30297 & collinear N\'eel AFM & -0.19171704 \\
219 & 0.325 & 0.3 & -0.29249 & 0.40258 & -0.30494 & inconclusive & -0.19952152 \\
220 & 0.325 & 0.325 & -0.26283 & 0.42891 & -0.30826 & inconclusive & -0.20813514 \\
221 & 0.325 & 0.35 & -0.23189 & 0.45511 & -0.31301 & hexagonal singlet & -0.21769514 \\
222 & 0.325 & 0.375 & -0.19938 & 0.48135 & -0.31927 & hexagonal singlet & -0.22825038 \\
223 & 0.325 & 0.4 & -0.165 & 0.50781 & -0.3272 & hexagonal singlet & -0.23983067 \\
224 & 0.325 & 0.425 & -0.12836 & 0.53468 & -0.33696 & hexagonal singlet & -0.25249784 \\
225 & 0.325 & 0.45 & -0.089039 & 0.56217 & -0.34879 & inconclusive & -0.26636156 \\
226 & 0.325 & 0.475 & -0.046476 & 0.59053 & -0.363 & canted $120^{\circ}$ & -0.28158349 \\
227 & 0.325 & 0.5 & 0 & 0.62004 & -0.37996 & canted $120^{\circ}$ & -0.29838405 \\
228 & 0.35 & 0 & -0.66246 & 0 & -0.33754 & FM & -0.20780748 \\
229 & 0.35 & 0.025 & -0.62901 & 0.049504 & -0.32149 & FM & -0.18506208 \\
230 & 0.35 & 0.05 & -0.59728 & 0.0946 & -0.30812 & FM & -0.16418495 \\
231 & 0.35 & 0.0625 & -0.58193 & 0.11575 & -0.30232 & FM & -0.15433384 \\
232 & 0.35 & 0.075 & -0.56687 & 0.13609 & -0.29704 & spin density wave & -0.15050528 \\
233 & 0.35 & 0.0875 & -0.55205 & 0.15569 & -0.29226 & collinear N\'eel AFM & -0.15370321 \\
234 & 0.35 & 0.1 & -0.53744 & 0.17463 & -0.28793 & collinear N\'eel AFM & -0.15721364 \\
235 & 0.35 & 0.125 & -0.50872 & 0.21072 & -0.28056 & collinear N\'eel AFM & -0.16416878 \\
236 & 0.35 & 0.15 & -0.48045 & 0.2448 & -0.27475 & collinear N\'eel AFM & -0.17098969 \\
237 & 0.35 & 0.175 & -0.45241 & 0.27724 & -0.27035 & collinear N\'eel AFM & -0.17770064 \\
238 & 0.35 & 0.2 & -0.42439 & 0.30833 & -0.26728 & collinear N\'eel AFM & -0.18435739 \\
239 & 0.35 & 0.225 & -0.39617 & 0.33836 & -0.26546 & collinear N\'eel AFM & -0.1910428 \\
240 & 0.35 & 0.25 & -0.36757 & 0.36757 & -0.26486 & collinear N\'eel AFM & -0.19787432 \\
241 & 0.35 & 0.275 & -0.33836 & 0.39617 & -0.26546 & collinear N\'eel AFM & -0.2050183 \\
242 & 0.35 & 0.3 & -0.30833 & 0.42439 & -0.26728 & collinear N\'eel AFM & -0.21270208 \\
243 & 0.35 & 0.325 & -0.27724 & 0.45241 & -0.27035 & inconclusive & -0.22119527 \\
244 & 0.35 & 0.35 & -0.2448 & 0.48045 & -0.27475 & inconclusive & -0.23073397 \\
245 & 0.35 & 0.375 & -0.21072 & 0.50872 & -0.28056 & hexagonal singlet & -0.24145318 \\
246 & 0.35 & 0.4 & -0.17463 & 0.53744 & -0.28793 & hexagonal singlet & -0.25341926 \\
247 & 0.35 & 0.425 & -0.13609 & 0.56687 & -0.29704 & hexagonal singlet & -0.26670775 \\
248 & 0.35 & 0.45 & -0.0946 & 0.59728 & -0.30812 & inconclusive & -0.28144769 \\
249 & 0.35 & 0.475 & -0.049504 & 0.62901 & -0.32149 & canted $120^{\circ}$ & -0.29783668 \\
250 & 0.35 & 0.5 & 0 & 0.66246 & -0.33754 & canted $120^{\circ}$ & -0.3161519 \\
251 & 0.375 & 0 & -0.70711 & 0 & -0.29289 & FM & -0.21338835 \\
252 & 0.375 & 0.025 & -0.66926 & 0.052672 & -0.27807 & FM & -0.18890517 \\
253 & 0.375 & 0.05 & -0.63381 & 0.10039 & -0.2658 & FM & -0.1665816 \\
254 & 0.375 & 0.0625 & -0.61681 & 0.12269 & -0.2605 & spin density wave & -0.16101557 \\
255 & 0.375 & 0.075 & -0.60022 & 0.1441 & -0.25568 & collinear N\'eel AFM & -0.16443912 \\
256 & 0.375 & 0.1 & -0.56804 & 0.18457 & -0.2474 & collinear N\'eel AFM & -0.17153429 \\
257 & 0.375 & 0.125 & -0.5369 & 0.22239 & -0.24071 & collinear N\'eel AFM & -0.1784772 \\
258 & 0.375 & 0.15 & -0.50648 & 0.25806 & -0.23545 & collinear N\'eel AFM & -0.18527213 \\
259 & 0.375 & 0.175 & -0.47651 & 0.292 & -0.23149 & collinear N\'eel AFM & -0.19195995 \\
260 & 0.375 & 0.2 & -0.44672 & 0.32456 & -0.22872 & collinear N\'eel AFM & -0.19860148 \\
261 & 0.375 & 0.225 & -0.41687 & 0.35604 & -0.22708 & collinear N\'eel AFM & -0.20527845 \\
262 & 0.375 & 0.25 & -0.38673 & 0.38673 & -0.22654 & collinear N\'eel AFM & -0.21210086 \\
263 & 0.375 & 0.275 & -0.35604 & 0.41687 & -0.22708 & collinear N\'eel AFM & -0.21922058 \\
264 & 0.375 & 0.3 & -0.32456 & 0.44672 & -0.22872 & collinear N\'eel AFM & -0.22684957 \\
265 & 0.375 & 0.325 & -0.292 & 0.47651 & -0.23149 & collinear N\'eel AFM & -0.23526742 \\
266 & 0.375 & 0.35 & -0.25806 & 0.50648 & -0.23545 & inconclusive & -0.24478084 \\
267 & 0.375 & 0.375 & -0.22239 & 0.5369 & -0.24071 & hexagonal singlet & -0.25563309 \\
268 & 0.375 & 0.4 & -0.18457 & 0.56804 & -0.2474 & hexagonal singlet & -0.26796668 \\
269 & 0.375 & 0.425 & -0.1441 & 0.60022 & -0.25568 & hexagonal singlet & -0.28189093 \\
270 & 0.375 & 0.45 & -0.10039 & 0.63381 & -0.2658 & inconclusive & -0.29756108 \\
271 & 0.375 & 0.475 & -0.052672 & 0.66926 & -0.27807 & inconclusive & -0.31521731 \\
272 & 0.375 & 0.5 & 0 & 0.70711 & -0.29289 & canted $120^{\circ}$ & -0.33520586 \\
273 & 0.4 & 0 & -0.75476 & 0 & -0.24524 & FM & -0.21934534 \\
274 & 0.4 & 0.025 & -0.71193 & 0.05603 & -0.23204 & FM & -0.19298027 \\
275 & 0.4 & 0.0375 & -0.69178 & 0.081878 & -0.22634 & FM & -0.18076845 \\
276 & 0.4 & 0.05 & -0.67233 & 0.10649 & -0.22118 & collinear N\'eel AFM & -0.1730302 \\
277 & 0.4 & 0.0625 & -0.65351 & 0.12999 & -0.2165 & collinear N\'eel AFM & -0.17652431 \\
278 & 0.4 & 0.075 & -0.63523 & 0.15251 & -0.21226 & collinear N\'eel AFM & -0.18007819 \\
279 & 0.4 & 0.1 & -0.60004 & 0.19496 & -0.205 & collinear N\'eel AFM & -0.18707018 \\
280 & 0.4 & 0.125 & -0.56628 & 0.23456 & -0.19916 & collinear N\'eel AFM & -0.19390279 \\
281 & 0.4 & 0.15 & -0.53356 & 0.27186 & -0.19457 & collinear N\'eel AFM & -0.20060559 \\
282 & 0.4 & 0.175 & -0.50154 & 0.30734 & -0.19112 & collinear N\'eel AFM & -0.20722659 \\
283 & 0.4 & 0.2 & -0.46989 & 0.34139 & -0.18872 & collinear N\'eel AFM & -0.21382901 \\
284 & 0.4 & 0.225 & -0.43833 & 0.37437 & -0.1873 & collinear N\'eel AFM & -0.2204934 \\
285 & 0.4 & 0.25 & -0.40659 & 0.40659 & -0.18683 & collinear N\'eel AFM & -0.22732423 \\
286 & 0.4 & 0.275 & -0.37437 & 0.43833 & -0.1873 & collinear N\'eel AFM & -0.23446208 \\
287 & 0.4 & 0.3 & -0.34139 & 0.46989 & -0.18872 & collinear N\'eel AFM & -0.24210241 \\
288 & 0.4 & 0.325 & -0.30734 & 0.50154 & -0.19112 & collinear N\'eel AFM & -0.25051576 \\
289 & 0.4 & 0.35 & -0.27186 & 0.53356 & -0.19457 & inconclusive & -0.26004291 \\
290 & 0.4 & 0.375 & -0.23456 & 0.56628 & -0.19916 & inconclusive & -0.27102665 \\
291 & 0.4 & 0.4 & -0.19496 & 0.60004 & -0.205 & hexagonal singlet & -0.2837221 \\
292 & 0.4 & 0.425 & -0.15251 & 0.63523 & -0.21226 & hexagonal singlet & -0.29830781 \\
293 & 0.4 & 0.45 & -0.10649 & 0.67233 & -0.22118 & hexagonal singlet & -0.31498145 \\
294 & 0.4 & 0.475 & -0.05603 & 0.71193 & -0.23204 & inconclusive & -0.33403625 \\
295 & 0.4 & 0.5 & 0 & 0.75476 & -0.24524 & canted $120^{\circ}$ & -0.35590354 \\
296 & 0.425 & 0 & -0.8064 & 0 & -0.1936 & FM & -0.22580005 \\
297 & 0.425 & 0.025 & -0.75785 & 0.059644 & -0.18251 & FM & -0.19736469 \\
298 & 0.425 & 0.0375 & -0.73523 & 0.087021 & -0.17775 & collinear N\'eel AFM & -0.18700696 \\
299 & 0.425 & 0.05 & -0.71354 & 0.11301 & -0.17344 & collinear N\'eel AFM & -0.19043632 \\
300 & 0.425 & 0.0625 & -0.69267 & 0.13778 & -0.16955 & collinear N\'eel AFM & -0.19389121 \\
301 & 0.425 & 0.075 & -0.6725 & 0.16145 & -0.16604 & collinear N\'eel AFM & -0.19731383 \\
302 & 0.425 & 0.1 & -0.63397 & 0.20599 & -0.16004 & collinear N\'eel AFM & -0.20405698 \\
303 & 0.425 & 0.125 & -0.59735 & 0.24743 & -0.15523 & collinear N\'eel AFM & -0.21067996 \\
304 & 0.425 & 0.15 & -0.56212 & 0.28642 & -0.15146 & collinear N\'eel AFM & -0.21721887 \\
305 & 0.425 & 0.175 & -0.52788 & 0.32348 & -0.14864 & collinear N\'eel AFM & -0.22372397 \\
306 & 0.425 & 0.2 & -0.49424 & 0.35909 & -0.14667 & collinear N\'eel AFM & -0.23025865 \\
307 & 0.425 & 0.225 & -0.46087 & 0.39362 & -0.14551 & collinear N\'eel AFM & -0.23690179 \\
308 & 0.425 & 0.25 & -0.42744 & 0.42744 & -0.14512 & collinear N\'eel AFM & -0.24375359 \\
309 & 0.425 & 0.275 & -0.39362 & 0.46087 & -0.14551 & collinear N\'eel AFM & -0.25094624 \\
310 & 0.425 & 0.3 & -0.35909 & 0.49424 & -0.14667 & collinear N\'eel AFM & -0.25866137 \\
311 & 0.425 & 0.325 & -0.32348 & 0.52788 & -0.14864 & collinear N\'eel AFM & -0.26715427 \\
312 & 0.425 & 0.35 & -0.28642 & 0.56212 & -0.15146 & collinear N\'eel AFM & -0.27677226 \\
313 & 0.425 & 0.375 & -0.24743 & 0.59735 & -0.15523 & inconclusive & -0.2879277 \\
314 & 0.425 & 0.4 & -0.20599 & 0.63397 & -0.16004 & hexagonal singlet & -0.30100377 \\
315 & 0.425 & 0.425 & -0.16145 & 0.6725 & -0.16604 & hexagonal singlet & -0.31629354 \\
316 & 0.425 & 0.45 & -0.11301 & 0.71354 & -0.17344 & hexagonal singlet & -0.33406868 \\
317 & 0.425 & 0.475 & -0.059644 & 0.75785 & -0.18251 & inconclusive & -0.35469399 \\
318 & 0.425 & 0.5 & 0 & 0.8064 & -0.1936 & inconclusive & -0.37870809 \\
319 & 0.45 & 0 & -0.86327 & 0 & -0.13673 & FM & -0.23290891 \\
320 & 0.45 & 0.0125 & -0.83487 & 0.032802 & -0.13233 & FM & -0.21705753 \\
321 & 0.45 & 0.025 & -0.80803 & 0.063593 & -0.12838 & collinear N\'eel AFM & -0.20366216 \\
322 & 0.45 & 0.0375 & -0.78257 & 0.092623 & -0.12481 & collinear N\'eel AFM & -0.20686217 \\
323 & 0.45 & 0.05 & -0.7583 & 0.1201 & -0.1216 & collinear N\'eel AFM & -0.21008276 \\
324 & 0.45 & 0.075 & -0.71278 & 0.17112 & -0.1161 & collinear N\'eel AFM & -0.21648424 \\
325 & 0.45 & 0.1 & -0.67049 & 0.21785 & -0.11166 & collinear N\'eel AFM & -0.22282644 \\
326 & 0.45 & 0.125 & -0.63066 & 0.26123 & -0.10812 & collinear N\'eel AFM & -0.22912595 \\
327 & 0.45 & 0.15 & -0.59267 & 0.30198 & -0.10535 & collinear N\'eel AFM & -0.23541717 \\
328 & 0.45 & 0.175 & -0.556 & 0.34072 & -0.10328 & collinear N\'eel AFM & -0.24174812 \\
329 & 0.45 & 0.2 & -0.52021 & 0.37795 & -0.10184 & collinear N\'eel AFM & -0.24817981 \\
330 & 0.45 & 0.225 & -0.48488 & 0.41413 & -0.101 & collinear N\'eel AFM & -0.25478842 \\
331 & 0.45 & 0.25 & -0.44964 & 0.44964 & -0.10072 & collinear N\'eel AFM & -0.26167049 \\
332 & 0.45 & 0.275 & -0.41413 & 0.48488 & -0.101 & collinear N\'eel AFM & -0.26895219 \\
333 & 0.45 & 0.3 & -0.37795 & 0.52021 & -0.10184 & collinear N\'eel AFM & -0.27680501 \\
334 & 0.45 & 0.325 & -0.34072 & 0.556 & -0.10328 & collinear N\'eel AFM & -0.28546984 \\
335 & 0.45 & 0.35 & -0.30198 & 0.59267 & -0.10535 & collinear N\'eel AFM & -0.29528656 \\
336 & 0.45 & 0.375 & -0.26123 & 0.63066 & -0.10812 & inconclusive & -0.30670387 \\
337 & 0.45 & 0.4 & -0.21785 & 0.67049 & -0.11166 & inconclusive & -0.32022119 \\
338 & 0.45 & 0.425 & -0.17112 & 0.71278 & -0.1161 & hexagonal singlet & -0.33628644 \\
339 & 0.45 & 0.45 & -0.1201 & 0.7583 & -0.1216 & hexagonal singlet & -0.35529539 \\
340 & 0.45 & 0.475 & -0.063593 & 0.80803 & -0.12838 & hexagonal singlet & -0.37771811 \\
341 & 0.45 & 0.5 & 0 & 0.86327 & -0.13673 & inconclusive & -0.40423363 \\
342 & 0.475 & 0 & -0.92704 & 0 & -0.07296 & FM & -0.24088004 \\
343 & 0.475 & 0.0125 & -0.89441 & 0.035142 & -0.070446 & collinear N\'eel AFM & -0.22409087 \\
344 & 0.475 & 0.025 & -0.86382 & 0.067984 & -0.068194 & collinear N\'eel AFM & -0.22681317 \\
345 & 0.475 & 0.05 & -0.80771 & 0.12793 & -0.064361 & collinear N\'eel AFM & -0.23239245 \\
346 & 0.475 & 0.075 & -0.75699 & 0.18174 & -0.061269 & collinear N\'eel AFM & -0.23807351 \\
347 & 0.475 & 0.1 & -0.71039 & 0.23082 & -0.058786 & collinear N\'eel AFM & -0.24383014 \\
348 & 0.475 & 0.125 & -0.66693 & 0.27625 & -0.056813 & collinear N\'eel AFM & -0.24966635 \\
349 & 0.475 & 0.15 & -0.62584 & 0.31888 & -0.05528 & collinear N\'eel AFM & -0.25560651 \\
350 & 0.475 & 0.175 & -0.58647 & 0.35939 & -0.054134 & collinear N\'eel AFM & -0.26169095 \\
351 & 0.475 & 0.2 & -0.5483 & 0.39836 & -0.053339 & collinear N\'eel AFM & -0.2679749 \\
352 & 0.475 & 0.225 & -0.51083 & 0.43629 & -0.052871 & collinear N\'eel AFM & -0.27453017 \\
353 & 0.475 & 0.25 & -0.47364 & 0.47364 & -0.052717 & collinear N\'eel AFM & -0.28144959 \\
354 & 0.475 & 0.275 & -0.43629 & 0.51083 & -0.052871 & collinear N\'eel AFM & -0.28885519 \\
355 & 0.475 & 0.3 & -0.39836 & 0.5483 & -0.053339 & collinear N\'eel AFM & -0.29691208 \\
356 & 0.475 & 0.325 & -0.35939 & 0.58647 & -0.054134 & collinear N\'eel AFM & -0.30585101 \\
357 & 0.475 & 0.35 & -0.31888 & 0.62584 & -0.05528 & collinear N\'eel AFM & -0.31600138 \\
358 & 0.475 & 0.375 & -0.27625 & 0.66693 & -0.056813 & collinear N\'eel AFM & -0.32782475 \\
359 & 0.475 & 0.4 & -0.23082 & 0.71039 & -0.058786 & inconclusive & -0.34190708 \\
360 & 0.475 & 0.425 & -0.18174 & 0.75699 & -0.061269 & hexagonal singlet & -0.35886992 \\
361 & 0.475 & 0.45 & -0.12793 & 0.80771 & -0.064361 & hexagonal singlet & -0.3792968 \\
362 & 0.475 & 0.475 & -0.067984 & 0.86382 & -0.068194 & hexagonal singlet & -0.40382146 \\
363 & 0.475 & 0.5 & 0 & 0.92704 & -0.07296 & hexagonal singlet & -0.43331629 \\
364 & 0.5 & 0 & -1 & 0 & 0 & isolated FM $J_t$-triangles & -0.25 \\
365 & 0.5 & 0.025 & -0.92704 & 0.07296 & 0 & collinear N\'eel AFM & -0.25381199 \\
366 & 0.5 & 0.05 & -0.86327 & 0.13673 & 0 & collinear N\'eel AFM & -0.258107 \\
367 & 0.5 & 0.075 & -0.8064 & 0.1936 & 0 & collinear N\'eel AFM & -0.26275356 \\
368 & 0.5 & 0.1 & -0.75476 & 0.24524 & 0 & collinear N\'eel AFM & -0.26768546 \\
369 & 0.5 & 0.125 & -0.70711 & 0.29289 & 0 & collinear N\'eel AFM & -0.27287761 \\
370 & 0.5 & 0.15 & -0.66246 & 0.33754 & 0 & collinear N\'eel AFM & -0.27833326 \\
371 & 0.5 & 0.175 & -0.62004 & 0.37996 & 0 & collinear N\'eel AFM & -0.28407755 \\
372 & 0.5 & 0.2 & -0.57919 & 0.42081 & 0 & collinear N\'eel AFM & -0.29015512 \\
373 & 0.5 & 0.225 & -0.53935 & 0.46065 & 0 & collinear N\'eel AFM & -0.29663074 \\
374 & 0.5 & 0.25 & -0.5 & 0.5 & 0 & collinear N\'eel AFM & -0.30359287 \\
375 & 0.5 & 0.275 & -0.46065 & 0.53935 & 0 & collinear N\'eel AFM & -0.3111608 \\
376 & 0.5 & 0.3 & -0.42081 & 0.57919 & 0 & collinear N\'eel AFM & -0.31949705 \\
377 & 0.5 & 0.325 & -0.37996 & 0.62004 & 0 & collinear N\'eel AFM & -0.32882804 \\
378 & 0.5 & 0.35 & -0.33754 & 0.66246 & 0 & collinear N\'eel AFM & -0.33947702 \\
379 & 0.5 & 0.375 & -0.29289 & 0.70711 & 0 & collinear N\'eel AFM & -0.35190908 \\
380 & 0.5 & 0.4 & -0.24524 & 0.75476 & 0 & inconclusive & -0.36676629 \\
381 & 0.5 & 0.425 & -0.1936 & 0.8064 & 0 & inconclusive & -0.38483518 \\
382 & 0.5 & 0.45 & -0.13673 & 0.86327 & 0 & hexagonal singlet & -0.40694848 \\
383 & 0.5 & 0.475 & -0.07296 & 0.92704 & 0 & hexagonal singlet & -0.43399488 \\
384 & 0.5 & 0.5 & 0 & 1 & 0 & isolated $J_h$-hexagons (=HS) & -0.46712927 \\

\end{longtable*}

\end{document}